\documentclass[runningheads,orivec]{llncs}
\usepackage[T1]{fontenc}

\usepackage{graphicx}
\usepackage{svg}

\usepackage{makecell}

\usepackage{enumitem}
\setlist[enumerate]{leftmargin=1.25cm}

\usepackage{pgf-pie}
\usepackage{xcolor}
\usepackage{adjustbox}

\usepackage{tabularx}
\usepackage{multirow}
\usepackage{wrapfig}
\usepackage{amsmath}

\DeclareMathOperator{\dom}{dom}
\DeclareMathOperator{\id}{id}
\usepackage{tikz}
\usetikzlibrary{shapes.geometric, arrows, calc, positioning}
\usetikzlibrary{shapes, arrows.meta, positioning, fit}

\usepackage{msc}

\usepackage{csquotes}

\usepackage[acronym]{glossaries}

\usepackage{listings}
\makeglossaries

\newacronym{cps}{CPS}{cyber–physical system}

\newacronym{ana}{ANA}{automotive network architecture}

\newacronym[
	longplural={Attack Resilience Hyperproperties}
]{arh}{ARH}{Attack Resilience Hyperproperty}

\newacronym{dy-m}{DY-model}{Dolev-Yao model}

\newacronym[
	longplural={threat and risk analyses}
]{tara}{TARA}{threat and risk analysis}

\newacronym{tcu}{TCU}{telematic control unit}

\newacronym{pki}{PKI}{public key infrastructure}

\newacronym{rq}{RQ}{research question}

\newacronym{ecu}{ECU}{electronic control unit}
\newacronym{ota}{OTA}{over-the-air}

\newacronym{fv}{FV}{formal verification}

\newacronym{impact}{ImpACT}{Improved Adversary Compromise Tool}
\newacronym{roadminer}{ROAD-Miner}{\enquote{Reconstructing Origins of Adversarial Damage}-Miner}

\newacronym[
	longplural={hyperproperties}
]{hp}{HP}{hyperproperty}

\newacronym[
	longplural={security properties}
]{sp}{SP}{security property}

\newacronym{crash-m}{CRASH-model}{\enquote{Compromised Realm Adversary System Hack} model}

\newacronym{dfa}{DFA}{deterministic finite-state automaton}

\newacronym{tamarin}{Tamarin}{the Tamarin Prover}

\newacronym{pm}{PM}{process mining}

\newacronym{dag}{DAG}{directed acyclic graph}

\newacronym{bms}{BMS}{battery management system}

\newacronym{dgw}{DGW}{domain gateway}

\newacronym{prom}{ProM}{Process Mining Toolkit}

\newacronym{idhm}{iDHM}{interactive Data-aware Heuristics Mine}

\newacronym{xes}{XES}{eXtensible Event Stream}

\newacronym{dfg}{DFG}{directly-follows graph}

\newacronym{exact}{ExACT}{Extended Adversary Compromise Tool}

\newacronym{adas}{ADAS}{advanced driver assistance system}

\usepackage{algorithm}
\usepackage[noEnd=true,indLines=false,spaceRequire=false]{algpseudocodex}
\usepackage{mathtools}
\usepackage{amsfonts}
\usepackage{MnSymbol}

\usepackage{subcaption}

\usepackage{bbding}

\begin{document}
	\title{Process-Mining of Hypertraces: Enabling Scalable Formal Security Verification of (Automotive) Network Architectures}
	\titlerunning{Process-Mining for Formal Security Verification of Automotive Architectures}
	\author{
		Julius Figge \Envelope\inst{1,2}
		\and
		David Knuplesch\inst{2}
		\and
		Andreas Maletti\inst{1}
		\and
		Dragan Zuvic\inst{2}
	}
	\authorrunning{J. Figge et al.}

	\institute{
		Institute of Computer Science, Leipzig University, 04109 Leipzig, Germany\\
		\email{\{julius.figge,andreas.maletti\}@uni-leipzig.de}\\
		\and
		Mercedes-Benz Tech Innovation GmbH, 89081 Ulm, Germany\\
		\email{\{julius.figge,david.knuplesch,dragan.zuvic\}@mercedes-benz.com}\\
	}
	\maketitle              %

	\begin{abstract}
		The automotive domain is transitioning: vehicles act as rolling servers, persistently connected to numerous external entities.
		This connectivity, combined with rising on-board computing power for \acrlongpl{adas} and similar use cases, creates escalating challenges for securing \acrlongpl{ana}.
		This work advances the security analysis of internet-connected \acrlongpl{ana} and their protocols.
		We introduce a strong, active adversary model tailored to the automotive domain.
		We substantially extend security protocol verification possible based on \acrfullpl{arh} by introducing a verification-orchestration algorithm.
		Furthermore, we provide methods for comparative attribution of \acrlong{sp} invalidations to specific, fine-grained component compromises.
		We present a novel integration of \acrlong{fv} and \acrlong{pm}.
		By utilizing \acrshort{arh} counterexample traces for \acrlong{pm}, we systematically identify and aggregate attacker behavior that causes \acrlong{sp} invalidations.
		This pipeline enables in-depth understanding of root causes and attack paths leading to protocol-security invalidations.
		We demonstrate real-world applicability through a prototype and case study on the secure transmission of \acrlong{bms} data within an automotive network architecture.
	\end{abstract}

	\section{Introduction}
		Automotive vehicles are undergoing a major transformation~\cite{lauserSecurityAnalysisAutomotive2020}, with the emergence of software-defined vehicles~\cite{asame.v.AsamSovdServiceOriented2022} and ubiquitous connectivity to external entities, e.g. mandatory over-the-air updates~\cite{seoFormallyVerifiedSoftware2023a}, becoming the norm.
		Vehicles are evolving into rolling datacenters with high-performance compute units using hypervisors and virtual machines~\cite{vamourSecurityOvertheAirSoftware2023,huelsewiesServerBasedArchitectureTransformation2021}, increasingly exposing them to broader cybersecurity challenges, including e.g.\ the shift to post-quantum cryptography~\cite{lohmillerSurveyPostQuantumCryptography2025} and required hardening against (remote) compromise.
		These challenges are compounded by regulation, including UN Regulation No. R156 (over-the-air updates), UN Regulation No. R155 (cybersecurity and software update management systems), and ISO/SAE 21434 (Road vehicles – Cybersecurity engineering)~\cite{seoFormallyVerifiedSoftware2023,marksteinerTARATestAutomated2023}.
		In parallel, vehicle functionality advances through systems such as \gls{adas} and battery-electric platforms, introducing further safety challenges that necessitate security.

		\subsubsection*{Motivation and Research Gap}
			Component compromise in the automotive domain is expected; its occurrence is a matter of when, not if~\cite{lampeIntrusionDetectionAutomotive2023,uptanestandardsgroupUptaneSecuringDelivery2021}.
			This motivates the comprehensive analysis of \glspl{ana} security under (partial) compromise as required by automotive regulations (e.g., ISO~21434)~\cite{marksteinerTARATestAutomated2023}.
			\Acrlong{fv} is well suited to prove the conditions under which \glspl{sp} of protocols in \glspl{ana} become invalid.\\
			Simultaneously, research gaps remain in \gls{fv} capabilities required to analyse \glspl{ana}.
			In particular, a comprehensive view is lacking, which would allow assessing the impact of different possible partial compromises.
			\emph{State-space explosion} limits comprehensive granularity, because of the trade-off between detail and analysis complexity.
			Thus, \gls{fv} focuses on refuting the examined \glspl{sp}~\cite{clarkeModelCheckingState2012}, yielding only isolated compromise scenarios as counterexamples.
			This e.g. hinders \emph{comparative analyses} of compromise impacts on \gls{sp} validity and the verification of \emph{fine-grained permission models}.\\
			We previously investigated these security challenges in \glspl{ana}~\cite{figgeApplicationsFormalVerification2024,figgeattackresiliencehyperproperties2026}.
			In particular, with \glspl{arh}~\cite{figgeattackresiliencehyperproperties2026}, we provide a systematic method to analyze protocol \glspl{sp} invalidation.
			Grounded in security-protocol verification, \glspl{arh} are \glspl{hp} that attribute \gls{sp}'s invalidation to protocol components (e.g., control units and network domains).
			For instance, they permit the identification of components that are necessary for the invalidation of a given \gls{sp}, but are not sufficient when considered in isolation.
			Despite these strengths, the above research gaps remain for \glspl{arh}, but some are addressable.
			However, their foundational approach imposes a conceptual limitation:
			\Gls{arh} analysis answers \enquote{\emph{who}} is responsible for invalidation but not \enquote{\emph{how}} adversarial behavior invalidates a protocol’s \glspl{sp}.
			These \gls{arh} limitations should be addressed by developing a method to synthesize comprehensive \glspl{sp} invalidating  attacker behaviors.
			This is especially relevant since \gls{hp}-based attribution obscures the underlying counterexample-traces, which renders the manual analysis of many traces, which is required to understand the possible attacker behavior, impractical.

		\subsubsection*{Research Questions}
			Our research questions \textbf{RQ1-RQ5} delineate this contribution's scope and limits and explicitly address the identified research gap.\\
			To reflect automotive architectural specifics and attacker's operating conditions, we must define a suitable attacker model (\textbf{RQ1}) that captures adversarial behavior, which is observed in the automotive domain, while not being limited to it.
			Current \gls{arh} analysis is constrained by state-space explosion.
			To advance the state of art and ensure practical applicability, we devise algorithmic improvements that reduce the number of scenarios to analyze (\textbf{RQ2}).
			With respect to the current capabilities of \gls{arh} analysis, it is desirable to enhance granularity by incorporating comparative assessments across \glspl{sp} (\textbf{RQ3}) and by distinguishing between different compromise modalities (\textbf{RQ4}).
			This increases both the level of detail and depth of the analysis.
			While current approaches are limited to compromise attribution without identifying the behavior that invalidates \glspl{sp}, our work focuses on comprehensive attacker-behavior identification and the development of necessary analysis techniques (\textbf{RQ5}).
			\begin{enumerate}[label=\textbf{RQ\arabic*},noitemsep, topsep=0.5em,leftmargin=*]
				\item How can an attacker model be extended to represent strong, active adversaries in \acrlongpl{ana} who participate maliciously in protocol flows?%
				\item Which approaches can make \acrshort{arh}-based \acrlong{hp} analysis scalable by reducing the number of compromise scenarios that must be verified?%
				\item Which methods enable the comparison of compromise scenarios across multiple \acrlongpl{sp} to assess and contrast invalidation conditions?%
				\item How can compromise modalities (e.g. read vs. write permissions) be modeled and analyzed to explain their impact on the invalidation of \acrlongpl{sp}?%
				\item How can \acrshort{arh} analysis be extended to identify and summarize adversarial behavior that causes invalidation of \acrlong{sp}?%
			\end{enumerate}

		\subsubsection*{Contributions}
			Our contributions \textbf{C1–C4}, detailed in the next section, directly address \textbf{RQ1–RQ5}.
			We map each contribution to its corresponding research question.
			\begin{enumerate}[label=\textbf{C\arabic*},noitemsep, topsep=0.5em,leftmargin=*]
				\item Adversarial \emph{\acrshort{crash-m}}
				\item Component compromise impact identification\begin{enumerate}[label=\textbf{C2.\arabic*},noitemsep, topsep=0pt,leftmargin=*]
					\item \acrshort{arh} verification orchestration algorithm
					\item Comparative multi-lemma analysis
					\item Fine granular adversarial permissions
				\end{enumerate}
				\item Comprehensive adversarial behavior analysis\begin{enumerate}[label=\textbf{C3.\arabic*},noitemsep, topsep=0pt,leftmargin=*]
					\item Interface between \acrshort{fv} and \acrlong{pm}
					\item Synthetic event-log generation
				\end{enumerate}
				\item Prototypical implementation \acrshort{impact} \& \acrshort{roadminer}
			\end{enumerate}
			Our \emph{\gls{crash-m}} (see Sect.~\ref{chap:adversary-model}), contribution~\textbf{C1}, successfully answers \textbf{RQ1} by providing an adversary model tailored to strong, active attackers in \glspl{ana} that covers permission levels, compromise of protocol entities and network segments, message extraction and injection, as well as complete takeover of entities including their full internal knowledge.\\
			Our second contribution~\textbf{C2} on \emph{identifying component-compromise impact} comprises three sub-contributions.
			For \textbf{C2.1} we present an \emph{\gls{arh} verification orchestration algorithm} that significantly reduces \gls{hp} verification runtime and complexity via algorithmic improvements and thus enables the analysis of large problem spaces and complex, fine-grained compromise scenarios as required by \textbf{RQ2}.
			For \textbf{C2.2} we provide an \emph{extended comparative analysis of \glspl{arh}} (see Sect.~\ref{chap:multiple-lemmas}) to answer \textbf{RQ3} on the impact of different compromise scenarios across \acrlongpl{sp}.
			With \textbf{C2.3} we provide \emph{fine-grained adversarial permissions} (see Sect.~\ref{chap:fine-granular-compromise}) to answer \textbf{RQ4} on analyzing how different compromise types and variants invalidate \acrlongpl{sp}.\\
			\textbf{C3} comprises of two sub-contributions that enable a \emph{comprehensive adversarial behavior analysis} (see Sect.~\ref{chap:adversarial-behavior-analysis}) and answer \textbf{RQ5}.
			We provide an \emph{interface between \gls{fv} and \gls{pm}} in~\textbf{C3.1} via a formalization of component definitions, translation elements, and mappings linking both fields.
			The \emph{synthetic event-log generation} \textbf{C3.2} enables the use of security protocol verification traces with \gls{pm} algorithms to analyze the adversarial behavior.
			At the same time, the behavioral analyses \textbf{C3.1} and \textbf{C3.2} add value to the compromise impact identification \textbf{C2.2} and \textbf{C2.3} and further answer \textbf{RQ3–RQ4}.\\
			We validate the real-world applicability of our contributions \textbf{C1–C3} as an answer to \textbf{RQ1–RQ5} with the help of our prototypical implementations \emph{\acrshort{impact}} and \emph{\acrshort{roadminer}} (\textbf{C4}; see Sect.~\ref{chap:prototype}).

		\subsubsection*{Structure}
			We introduce our case study in Sect.~\ref{chap:case-study}, define the attacker model, present the \gls{ana} and protocol, and define the relevant \acrlongpl{sp}.
			We then present our contributions for identifying the impact of component compromises in Sect.~\ref{chap:identifying-impact-component-compromise}, a fine-grained subdivision of compromises in Sect.~\ref{chap:fine-granular-compromise}, comparative \gls{sp} analysis in Sect.~\ref{chap:multiple-lemmas}, and an algorithmic verification-orchestration approach in Sect.~\ref{chap:arh-verification-orchetration-algorithm} that enables \gls{arh} verification under our attacker model.
			Next, we present our contribution for the comprehensive adversarial-behavior analysis in Sect.~\ref{chap:adversarial-behavior-analysis}.
			We validate real-world relevance by applying the case study to our prototypical implementation in Sect.~\ref{chap:prototype}.
			Finally, we discuss related work in Sect.~\ref{chap:related-work} and highlight how our contributions differs from existing research.
			We conclude with a summary and outlook in Sect.~\ref{chap:conclusion-and-outlook}.

	\section{Case Study}\label{chap:case-study}
	In the following, we describe our case study using a fictional \gls{bms} protocol.
	We first present our contribution of the \gls{crash-m} as the adversarial model and subsequently the exemplary \gls{ana}.\\
	We provide a level of detail that reflects an industrially relevant, real-world example while remaining comprehensible.

	\subsection{Adversarial Model}\label{chap:adversary-model}
		We present our attacker model as an enhancement of the \gls{dy-m}, the standard adversarial model used in verification tools such as \gls{tamarin}.
		Our \acrfull{crash-m} (see Fig.~\ref{fig:crash-adversary-model}) is designed with a focus on \glspl{ana} and \acrlongpl{cps}.
		It adds extensions and modifications to better suit these areas than the \gls{dy-m}.

		\begin{wrapfigure}{r}{0.65\textwidth}
			\vspace*{-1em}
			\centering
			\resizebox{0.65\textwidth}{!}{
				\begin{tikzpicture}
					\draw[thick] (-0.5,0) rectangle (3,2);
					\draw[thick] (0.5,0) -- (0.5,2);
					\draw[thick] (2,0) -- (2,2);
					\draw[dashed, thick] (-0.625,-0.125) rectangle (3.125,2.125);
					\node (A) at (0,1) {\textbf{A}};
					\node (B) at (1.25,1.5) {\textbf{B}};
					\node (C) at (1.25,0.5) {\textbf{C}};
					\node (D) at (2.55,1) {\textbf{D}};
					\draw[blue,->] (A) -- node[pos=0.05, above] {$m_1$} (B);
					\draw[blue,->] (B) -- node[pos=0.2, right] {$m_2$} (C);
					\draw[blue,->] (C) -- node[pos=0.95, below] {$m_3$} (D);

					\draw[thick] (4,0) rectangle (5,2);
					\node (E) at (4.5,1) {\textbf{E}};
					\draw[blue,->] (D) -- node[pos=0.5, above] {$m_4$} (E);

					\draw[thick] (6,0) rectangle (7.5,2);
					\draw[dashed, thick] (5.875,-0.125) rectangle (7.625,2.125);
					\node (F) at (6.75,1) {\textbf{F}};
					\node at (6.75,0.7) {$k_1, k_2$};
					\draw[thick] (6.75,1) circle (0.65);
					\draw[blue,->] (E) -- node[pos=0.35, above] {$m_4$} (F);

					\node (AD) at (9,1) {\textbf{AD}};

					\draw[red, dashed, thick] (0.85,0.2) rectangle (1.8,1.8);
					\draw[purple, dashed, thick] (0.7,0.1) rectangle (1.95,1.9);
					\draw[red, -{Latex[scale=1.2]}] (1.7,1.8) to[out=20, in=140] node[pos=0.45, below] {$m_2$} (AD);
					\draw[purple, -{Latex[scale=1.2]}] (AD) to[out=-140, in=-20] node[pos=0.55, above] {$m_2'$} (1.9,0.1);
					\draw[red, dashed, thick] (4.5,1) circle (0.2125);
					\draw[purple, dashed, thick] (4.5,1) circle (0.3);
					\draw[red, -{Latex[scale=1.2]}] (4.5,1.25) to[out=20, in=140] node[pos=0.2, above] {$m_4$} (AD);
					\draw[purple, -{Latex[scale=1.2]}] (AD) to[out=-140, in=-20] node[pos=0.8, below] {$m_4'$} (4.5,0.7);
					\draw[red, dashed, thick] (6.75,1) circle (0.55);
					\draw[purple, dashed, thick] (6.75,1) circle (0.6);
					\draw[red, -{Latex[scale=1.2]}] (6.75,1.55) to[out=20, in=140] node[pos=0.65, below] {\textit{$k_1$}} (AD);
					\draw[purple, -{Latex[scale=1.2]}] (AD) to[out=-140, in=-20] node[pos=0.35, above] {\textit{$k_1'$}} (6.8,0.4);
				\end{tikzpicture}
			}
			\vspace{-2.25em}
			\captionof{figure}{\acrshort{crash-m}}
			\label{fig:crash-adversary-model}
			\vspace{-2.25em}
		\end{wrapfigure}
		At its core, the \gls{crash-m} introduces two key changes.
		These changes focus on entity and domain compromise  and align with our threat model, which considers physical access to networks (including bus systems and Ethernet subnets).
		Access can lead to takeover and compromise of entities such as \glspl{ecu}.\\
		First, the attacker model supports network segmentation (e.g. into domains).
		An attacker can interact with traffic (messages on the network) only after compromising a network segment (domain) or a participant (entity).
		It can intercept messages inside compromised segments (e.g. $m_2$ in Fig.~\ref{fig:crash-adversary-model}) and inject forged messages ($m_2'$).
		Alternatively, it can intercept messages to or from compromised entities ($m_4$) or inject such messages ($m_4'$).\footnote{Extended capabilities, such as suppressing or modifying messages, are possible but unnecessary here and are excluded from our model.}
		This first extension thus refines the existing \gls{dy-m} capabilities, while the second extension grants the attacker completely new abilities.
		It can also read and even alter the internal knowledge ($k_1$) for all compromised entities.

	\subsection*{Automotive Network Architecture}
		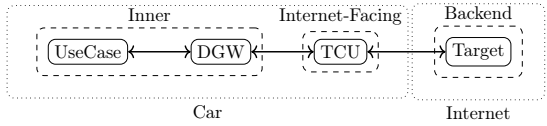
\begin{wrapfigure}{l}{0.6\textwidth}
			\vspace{-3.25em}
			\begin{flushleft}
				\resizebox{0.6\textwidth}{!}{
					\begin{tikzpicture}[
						node distance=1.2cm,
						every node/.style={draw, rounded corners},
					bidirectional/.style={<->, thick}
						]

						\node (UseCase) {UseCase};
						\node (DGW) [right=of UseCase] {DGW};
						\node (TCU) [right=of DGW] {TCU};
						\node (Target) [right=1.5cm of TCU] {Target};

						\node (Inner)[draw, dashed, fit=(UseCase) (DGW), label=above:Inner, inner xsep=6pt, inner ysep=6pt] {};
						\node (Internet-Facing)[draw, dashed, fit=(TCU), label=above:Internet-Facing, inner xsep=6pt, inner ysep=4pt] {};

						\node [draw, dotted, fit=(Inner) (Internet-Facing), label=below:Car, inner xsep=16pt, inner ysep=12pt] {};

						\node (Internet)[draw, dashed, fit=(Target), label=above:Backend, inner xsep=6pt, inner ysep=6pt] {};

						\node [draw, dotted, fit=(Internet), label=below:Internet, inner xsep=12pt, inner ysep=12pt] {};

						\draw[bidirectional] (UseCase) -- (DGW);
						\draw[bidirectional] (DGW) -- (TCU);
						\draw[bidirectional] (TCU) -- (Target);

					\end{tikzpicture}
				}
				\vspace{-1.75em}
				\caption{Exemplary \acrshort{ana}}
				\label{fig:exemplary-usecase-architecture}
			\end{flushleft}
			\vspace{-3em}
		\end{wrapfigure}
		Our exemplary \gls{ana} (see Fig.~\ref{fig:exemplary-usecase-architecture}) is a simplified abstraction of a real-world \gls{ana}.
		The \gls{ana} consists of two main components: the vehicle and the Internet.
		The vehicle is represented by two network segments with different levels of criticality, regarding security and safety.
		On the left is the \emph{Inner} domain, the segment of higher criticality, hosting the \emph{UseCase} \gls{ecu}.
		This placeholder is the protocol's first communication partner.
		Examples of highly critical \glspl{ecu} include components such as \gls{bms} and autonomous driving systems.
		The \emph{\acrfull{dgw}} bridges the \emph{Inner} and \emph{Internet-Facing} domains and communicates with the UseCase.
		In the less critical \emph{Internet-Facing} domain, the \emph{\gls{tcu}} is the communication partner, connecting the vehicle to the Internet.\\
		On the Internet, the (vehicle manufacturer's) \emph{Backend} segment hosts the exemplary \emph{Target} service, which is the protocol's recipient and communication partner for the \gls{tcu}.

	\subsection*{Protocol Description}
		Finally, we present our example protocol, based on the previously presented \gls{ana} and used in the evaluation of the prototypical implementation (see Sect.~\ref{chap:prototype}).
		The protocol is intentionally insecure to facilitate demonstration.\\
		It targets message authenticity\footnote{From the perspective of the target, i.e. receiver, service.}~\cite{loweHierarchyAuthenticationSpecifications1997a}, while message confidentiality relies on closed network segments and communication channels.\\
		We motivate the example with a \gls{bms} protocol, where an electric vehicle transmits battery charge status, charge cycles, and temperature to the manufacturer to optimize battery lifespan.
		The protocol below outlines \gls{bms} data transmission in concise, abstracted form.\\
		\begin{wrapfigure}{r}{0.6\textwidth}
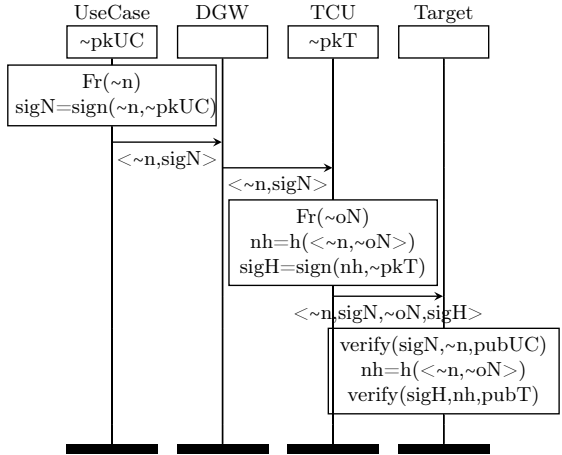

			\vspace*{-5.5em}
			\begin{flushleft}
				\resizebox{0.6\columnwidth}{!}{
					\hspace*{-0.5cm}
					\begin{msc}[small values, msc keyword=, instance width=0.5cm, first level height=0.1cm,level height=0.1cm,instance width=1.4cm, left inline overlap=1.7cm, right inline overlap=1.6cm]{}
						\drawframe{no}

						\declinst{UseCase}{UseCase}{$\sim$pkUC}
						\declinst{DGW}{DGW}{}
						\declinst{TCU}{TCU}{$\sim$pkT}
						\declinst{Target}{Target}{}

						\action*{%
							\parbox{3cm}%
							{\centering Fr($\sim$n)
								sigN=sign($\sim$n,$\sim$pkUC)
							}}
						{UseCase}
						\nextlevel[12]

						\mess[label position=below]{<$\sim$n,sigN>}{UseCase}{DGW}
						\nextlevel[4]

						\mess[label position=below]{<$\sim$n,sigN>}{DGW}{TCU}
						\nextlevel[5]

						\action*{%
							\parbox{3cm}%
							{\centering Fr($\sim$oN)
								nh=h(<$\sim$n,$\sim$oN>)
								sigH=sign(nh,$\sim$pkT)
							}}
						{TCU}
						\nextlevel[15]

						\mess[label position=below]{<$\sim$n,sigN,$\sim$oN,sigH>}{TCU}{Target}
						\nextlevel[5]

						\action*{%
							\parbox{3.35cm}%
							{\centering
							verify(sigN,$\sim$n,pubUC)\\
							nh=h(<$\sim$n,$\sim$oN>)\\
							verify(sigH,nh,pubT)\\
							}}
						{Target}
						\nextlevel[15]
					\end{msc}
				}
			\end{flushleft}
			\vspace*{-2.25em}
			\caption{\acrshort{bms} Synchronization Protocol} \label{fig:exemplary-protocol-bms}
			\vspace*{-2.25em}
		\end{wrapfigure}
		The \gls{bms} $UseCase$ monitors relevant battery parameters to be transmitted to the backend and stores them as data~${\sim}n$.
		The data~${\sim}n$ is signed by the \gls{bms} with its private key~${\sim}pkUC$ and sent to the $TCU$ via the $DGW$.
		The $TCU$ adds vehicle-specific information~${\sim}\!oN$ and signs a hash of the combined data~$\langle {\sim}n,{\sim}oN \rangle$ with its private key~${\sim}pkT$.
		The data is then transmitted to the manufacturer's $Backend$ and the corresponding \gls{bms}-Service $Target$, which validates the \gls{bms} data and vehicular information integrity~$verify(\dots)$, concluding the protocol.

	\subsection*{Security Properties}\label{chap:security-properties}
		We define the protocol's \acrfullpl{sp} under the specified threat model (see Sect. \ref{chap:adversary-model}), including resistance to an active attacker and partial compromise scenarios.

		\subsubsection*{Secrecy of transmitted data}
			The secrecy property requires that for every nonce (${\sim}n, {\sim}oN$) in any protocol execution, if at no time an entity's private key is revealed from the \gls{pki}, then there exists no time point at which an adversary knows the nonces.
			Consequently, \gls{bms} and vehicular data must remain unknown to the adversary, if no private key is leaked.

		\subsubsection*{Authenticity of UseCase messages}
			The authenticity property\footnote{See List. \ref{list:tamarin-lemma-authenticity}.} requires that for any protocol execution ending with the target receiving the nonces, if at no prior time point an entity's private key was revealed, there must exist an earlier time point at which the $UseCase$ initiated the protocol run with the same nonce~${\sim}n$.
			In other words, the \gls{bms} data received by the target must have been sent (and generated) by the $UseCase$.

	\section{Identifying Impact of Component Compromise}\label{chap:identifying-impact-component-compromise}
	We introduce our approach for the impact identification of component compromise through \acrlong{arh} analysis, as an essential foundation for the following contributions.
	This enables analysis of complex compromise scenarios with fine-grained permission models and multiple \acrfullpl{sp} across all possible respective combinations.

	\subsection{Approach Formalization}
		We subsequently formalize the components necessary for \gls{arh} analysis.
		This formalization builds on our prior work on \glspl{arh}~\cite{figgeattackresiliencehyperproperties2026}.
		We first define \glspl{ana}, comprising the architecture and its application protocols.
		We then extend this with domain-specific \gls{crash-m} attacker model elements, \glspl{sp}, and a model of protocol executions, including component states, compromise scenarios and adversarial behavior.\\
		An \emph{\gls{ana}} is a directed graph~$A = (E, L, D)$, with nodes~$E$ representing \emph{entities} such as \glspl{ecu} or services, edges~$L \subseteq E \times E$ as directed communication \emph{links}, and the partition~$D$ of $E$, representing \emph{domains}, i.e. network segments.
		For each~$e \in E$, we require~$(e,e) \in L$, providing self-links for modeling internal computations.
		For each domain~$d \in D$, a special node~$e_d \in E$ without links is included to support compromise modeling.\\
		A \emph{message}~$m = (l, n) \in L \times N$ is a tuple with link~$l \in L$ (communication partners) and contents~$n \in N$\footnote{Typically a nonce from a set~$N$ of random numbers.}, forming the finite set of all messages~$M$.\\
		An entity's~$e \in E$ internal computations, such as cryptographic operations on messages, are represented by internal messages~$((e, e), n)$, enabling new values and stored knowledge from a finite set~$K$.
		To model dynamic knowledge possession, we use a \gls{dfa}~\cite{yuRegularLanguages1997} with state set~$Q = S \times \mathcal{P}(K)$ for finite state set~$S$.\\
		The state's components~$(s, K') \in Q$ represent an entity's internal state~$s \in S$ and current knowledge~$K' \subseteq K$.
		The \gls{dfa} is a tuple~$(Q, \mathcal{A}, q_0, \delta, F)$ with~$Q = S \times \mathcal{P}(K)$ for finite~$S$, \emph{action} alphabet~$\mathcal{A} \subseteq M$, initial state~$q_0 = (s_0, k_0) \in Q$ with initial state~$s_0$ and knowledge~$k_0$, transition function~$\delta \colon Q \times \mathcal{A} \to Q$, and set of final states~$F \subseteq Q$.
		The transition function~$\delta$ extends to~$\Delta \colon Q \times \mathcal{A}^* \to Q$ via~$\Delta(q, \varepsilon) = q$ and~$\Delta(q, aw) = \Delta(\delta(q, a), w)$ for all~$q \in Q$, $a \in \mathcal{A}$, and~$w \in \mathcal{A}^*$.
		Final states~$F$ are unused.\\
		Each entity $e \in E$ is associated with a \gls{dfa}~$T(e)$, modeling its dynamic behavior via message contents; automata are not used for domains~$d \in E$.
		Transitions in~$T(e) = (Q, \mathcal{A}, q_0, \delta, F)$ occur only on messages involving the entity.
		We assume~$\mathcal{A} = M$ and~$\delta(q,a) = q$ for any message~$a \in \mathcal{A}$ not involving entity~$e$.
		Thus, the state of~$T(e)$ remains unchanged for messages where~$e$ is neither sender nor receiver.
		For convenience, we assume a special message content~$\mathrm{tick} = 0$, which, when sent as~$(l, \mathrm{tick})$ with~$l \in L$, causes no reaction, i.e. it does not change any entity's knowledge or state.\\
		A \emph{protocol} is a finite message sequence~$w \in M^*$.
		Every entity~$e \in E$ reacts as described below; domain nodes~$e_d$ with~$d \in D$ are irrelevant because they lack links, i.e.\ not occur in regular protocol execution.
		Let~$T(e) = (Q, M, q_0, \delta, F)$ be the \gls{dfa} of~$e$, with knowledge after protocol execution~$w$ given by~$K(e, w) = K'$, where~$\Delta(q_0, w) = (s, K')$.
		A protocol~$w$ is \emph{valid} if for every prefix~$w' ((s, r), n)$ with final message~$((s, r), n)$, we have~$n \in K(s, w')$.\\
		We define the set of \emph{attacker capabilities}, i.e. possible permission combinations, as~$P = \{ \emptyset, \mathtt{r}, \mathtt{w}, \mathtt{rw} \}$\footnote{Extending the utilized permissions to e.g. execution $\mathtt{x}$ and deletion $\mathtt{d}$ is conceivable.}.
		Recall that $E$~denotes the components (entities or domains).
		A \emph{compromise} is a mapping~$c\, \colon E \to P$.
		In other words, a compromise assigns attacker capabilities to each component.
		For convenience, we denote the compromised components as $\dom(c) = \{\, e \in E \mid c(e) \neq \emptyset \,\}$.
		We use our \gls{crash-m} (see Fig.~\ref{fig:crash-adversary-model}).
		For any entity~$e \in E$ compromised with read permission $\mathtt{r}$~or~$\mathtt{rw}$, all messages to or from~$e$ can be intercepted.
		Similarly, messages~$((s, r), n)$ involving a read-compromised domain~$d \in D$ can be intercepted, i.e.~$c(e_d) \in \{\mathtt{r}, \mathtt{rw}\}$ and~$\{s,r\} \cap d \neq \emptyset$.
		The adversary also has full access to the knowledge of read-compromised entities.
		In other words, any message content~$n$ sent to or from a read-compromised entity or domain is immediately known to the adversary, along with the knowledge and internal state of read-compromised entities.
		Additionally, the adversary can inject messages to and from write-compromised entities~$c(e) \in \{\mathtt{w}, \mathtt{rw}\}$ and domains~$c(e_d) \in \{\mathtt{w}, \mathtt{rw}\}$, on behalf of the compromised components.
		For such messages, we mark the spoofed endpoint with a tick, yielding~$((s', r), n)$ or~$((s, r'), n)$, respectively.
		We use this notation to distinguish cases: non-compromised entities are oblivious to spoofing and react normally, including state and knowledge updates.
		Compromised entities react only to messages in which they are not marked.
		Otherwise, all entities, including compromised ones, react to protocol messages normally.
		This keeps the formalization simple while preserving strong adversarial behavior and capabilities.
		The adversary may use any possessed knowledge to craft injected messages; it is not limited to the spoofed component's knowledge.\\
		We define an \emph{Execution Trace} as an even-length message sequence \linebreak$w = m_1m'_1m_2m'_2 \cdots m_nm'_n \in M^*$, where messages~$m'_1, \dotsc, m'_n$ are controlled and injected by the adversary.
		The special message~$((e, e), \mathrm{tick})$ for any compromised entity~$e \in \dom(c)$ lets the adversary refrain from injecting a real message, modeling no reaction or change at that time.\\
		An execution trace~$w$ is \emph{valid} for compromise~$c$ if
		\[
			n \in
			\begin{cases}
				K(s, w') & \text{if } s \not\in \dom(c) \\
				\bigcup_{e \in E \colon c(e) \in \{\mathtt{r}, \mathtt{rw}\}} K(e, w') &  \text{otherwise}
			\end{cases}
		\]
		for every prefix~$w'((s, r), n)$ of~$w$.\\
		We define the set of possible \glspl{sp} as~$\mathtt{S}$.\\
		For a compromise~$c$, \acrlong{sp}~$S \in \mathtt{S}$, and an execution trace~$w$ valid for~$c$, we write~$w \models_c S$ if~$S$ holds for~$w$.
		For a set~$L$ of execution traces valid for~$c$, we define~$L \models_c S$ to be true iff~$w \models_c S$ for all~$w \in L$.
		For any language~$L \subseteq M^*$, we define~$C(S, L) = \{\, c\, \colon E \to P \mid \exists\, w \in L \text{ valid for } c \text{ with } w \not\models_c S \,\}$ as the set of compromises for which some valid~$w \in L$ violates \acrlong{sp}~$S$.
		The complement~$\overline{C(S, L)} = P^E \setminus C(S, L)$ represents compromises, where violations of the \acrlong{sp}~$S$ do not occur.

	\subsection{ARH Evaluation and Extension}
		We use our formalization to analyze \glspl{arh}, which enable examining responsibility and involvement in \gls{sp} invalidation across compromise scenarios.
		In other words, \glspl{arh} identify which components are responsible for or involved in an invalidation.
		To this end, we need to introduce a partial order on the capabilities in the expected manner $\emptyset < \mathtt{r} < \mathtt{rw}$ as well as $\emptyset < \mathtt{w} < \mathtt{rw}$, but $\mathtt{r} \not< \mathtt{w} \not< \mathtt{r}$.\footnote{This is the expected Boolean algebra for the 2 capabilities of read~$\mathtt{r}$ and write~$\mathtt{w}$.}
		For compromises $c, c' \colon E \to P$ we write~$c \preceq c'$ if $c(e) \leq c'(e)$ for all~$e \in E$.
		Finally, we define $c' - c\, \colon E \to P$ by $(c' - c)(e) = \emptyset$ for all~$e \in \dom(c)$ and $(c' - c)(e) = c'(e)$ otherwise.

		\subsubsection*{\acrshort{arh}}\label{chap:arh}
			For the analysis, we employ the following \glspl{arh}~\cite{figgeattackresiliencehyperproperties2026}:

			\begin{description}
				\item[\textbf{Necessary but not sufficient (NBNS)}] identifies individual components in multistep compromises of \gls{sp}~$S \in \mathtt{S}$, defined as \\
				$\text{NBNS} = \min_{\mathord{\preceq}} \{c \in \overline{C(S, L)} \mid  \exists c' \in C(S,L) \colon c \preceq c', c' - c \in \overline{C(S, L)} \}$.
				\item[\textbf{Never responsible for Compromise (NRFC)}] comprises scenarios that \\never contribute to invalidating a \gls{sp}, defined as \\
				$\text{NRFC}= \{c \in \overline{C(S,L)} \mid \forall c' \in C(S, L) \colon c \preceq c' \text{ implies } c' - c \in C(S, L) \}$.
				\item[\textbf{Minimal Compromise Scenario (MCS)}] captures the smallest compromises that invalidate a given \gls{sp}, defined as $\text{MCS} = \min_{\mathord{\preceq}} C(S, L)$.
				\item[\textbf{Single Point of Failure (SPOF)}] is a special case of MCS, defined as\\
				$\text{SPOF} = \{ c \in C(S, L) \mid \lvert \dom(c) \rvert = 1\}$.
			\end{description}

		\subsubsection*{Multiple \acrlongpl{sp}}\label{chap:multiple-lemmas}
			We extend the analysis beyond individual \gls{arh} by evaluating multiple \glspl{sp} simultaneously.
			This enables comparative analysis of \glspl{sp} by examining the compromise scenarios and execution traces that lead to their invalidation.
			Consequently, we can identify weaknesses and \glspl{arh} that span multiple \glspl{sp}.
			To that end, we define the set~$\mathbb{S}(c, L) \subseteq \mathtt{S}$ of \glspl{sp} that a given compromise scenario~$c$ causes to be invalidated by traces~$w \in L$ as
			\[
				\mathbb{S}(c, L) = \{ S \in \mathtt{S} \mid \exists w \in L \text{ valid for } c \text{ with } w \not\models_c S \}
			\]

		\subsubsection*{Fine Granular Compromise}\label{chap:fine-granular-compromise}
			Our proposed adversary \gls{crash-m} (see Sect. \ref{chap:adversary-model}) permits a wide range of attacker actions.
			We categorize these actions and propose \emph{fine-granular compromises} to model attacker capabilities.
			Permissions follow a decomposition into non-empty combinations of \emph{read}, \emph{write} access~$P = \{ \emptyset, \mathtt{r}, \mathtt{w}, \mathtt{rw} \}$.
			We provide the mapping~$c\, \colon E \to P$ between components, i.e. protocol entities or domains and respective permissions granted through compromise.
			For our example (see Fig. \ref{fig:exemplary-usecase-architecture}), an attacker may have write access to the Internet domain and only read access to the \gls{tcu}.
			Arbitrary per-component combinations are allowed; granting all capabilities recovers the previous coarse-grained model.
			This model better reflects scenario-dependent attacker capabilities (e.g. read-only memory extraction or physical man-in-the-middle injection) and supports verification workflows that identify the minimal permissions required to invalidate a \gls{sp} and isolate irrelevant permissions.

	\subsection{\acrshort{arh}-Verification Orchestration Algorithm}\label{chap:arh-verification-orchetration-algorithm}
		Introducing fine-grained attacker capabilities and multiple simultaneous \glspl{sp} significantly increases the number of scenarios for verification.
		\Gls{arh} verification must consider every compromise~$c\, \colon E \to P$ and each security property~$S \in \mathtt{S}$.
		Consequently, there are $\lvert P\rvert^{\lvert E\rvert} \cdot \lvert \mathtt{S} \rvert$ scenarios to verify.
		Memory demands are limiting the number of parallel executions of separate verifications.
		However, the time complexity grows exponentially in the number~$\lvert E\rvert$ of components and the number~$p$ of capabilities as $\lvert P \rvert = 2^p$.
		In our example we have $p = 2$ and thus~$\lvert P\rvert = 4$.
		On the  other hand, the time complexity grows only linearly in the number~$\lvert \mathtt{S} \rvert$ of \glspl{sp}, which renders this factor manageable for complexity considerations.
		Overall, we obtain long runtime and substantial computational demand.
		\begin{wrapfigure}{l}{0.575\textwidth}
			\centering
			\vspace{-2em}
			\resizebox{0.575\textwidth}{!}{%
				\begin{tikzpicture}[
					validated/.style={draw, circle, minimum size=2cm, fill=green!20},
					invalidated/.style={draw, circle, minimum size=2cm, fill=red!20},
					halfInvalidated/.style={draw, circle, minimum size=1cm},
					markedInvalidated/.style={draw, circle, minimum size=2cm, fill=red!10, dotted, thick, draw=red!60!black},
					edge/.style={thick, -{stealth[scale=1.7]}, draw=black!50, shorten >=2pt, shorten <=2pt},
					redhalfInvalidatedDottedEdge/.style={draw=red!70, -{stealth[scale=1.7]}, shorten >=2pt, shorten <=2pt},
					redEdge/.style={thick, dotted, draw=red!70!black!70, -{stealth[scale=1.7]}, shorten >=2pt, shorten <=2pt},
				]

						\node[validated] (none) at (0,-3.25) {$\bot$};

						\node[validated] (Ar) at (-5.75,-1) {$A \mapsto \text{r}$};
						\begin{scope}
							\clip (-3.5,-0.25) circle (1cm);
							\fill[green!20] (-3.5,-0.25) circle (1cm);
						\end{scope}
						\begin{scope}
							\clip (-3.5,-0.25) circle (1cm);
							\clip (-3.5,-0.25) -- ++(1,0) arc[start angle=0,end angle=180,radius=1cm] -- cycle;
							\fill[red!20] (-3.5,-0.25) circle (1cm);
						\end{scope}
						\node[halfInvalidated] (Arw) at (-3.5,-0.25) {$A \mapsto \text{rw}$};
						\node[validated] (Aw) at (-1.25,-1) {$A \mapsto \text{w}$};

						\node[validated] (Br) at (1.25,-1) {$B \mapsto \text{r}$};
						\node[markedInvalidated] (Brw) at (3.5,-0.25) {$B \mapsto \text{rw}$};
						\node[invalidated] (Bw) at (5.75,-1) {$B \mapsto \text{w}$};

						\node[validated] (ArBr) at (-7,2.5) {$\begin{array}{c}A \mapsto \text{r} \\ B \mapsto \text{r}\end{array}$};
						\node[markedInvalidated] (ArBw) at (-5.5,4) {$\begin{array}{c} A \mapsto \text{r} \\ B \mapsto \text{w} \end{array}$};
						\node[markedInvalidated] (ArBrw) at (-4,5.5) {$\begin{array}{c} A \mapsto \text{r} \\ B \mapsto \text{rw}\end{array}$};

						\begin{scope}
							\clip (-2.5,7.5) circle (1.1cm);
							\fill[green!20] (-2.5,7.5) circle (1.1cm);
						\end{scope}
						\begin{scope}
							\clip (-2.5,7.5) circle (1.1cm);
							\clip (-2.5,7.5) -- ++(1.1,0) arc[start angle=0,end angle=180,radius=1.1cm] -- cycle;
							\fill[red!20] (-2.5,7.5) circle (1.1cm);
						\end{scope}
						\node[halfInvalidated] (ArwBr) at (-2.5,7.5) {$\begin{array}{c} A \mapsto \text{rw} \\ B \mapsto \text{r}\end{array}$};
						\node[markedInvalidated] (ArwBw) at (0,7.5) {$\begin{array}{c} A \mapsto \text{rw} \\ B \mapsto \text{w}\end{array}$};
						\node[markedInvalidated] (ArwBrw) at (2.5,7.5) {$\begin{array}{c} A \mapsto \text{rw} \\ B \mapsto \text{rw}\end{array}$};

						\node[validated] (AwBr) at (4,5.5) {$\begin{array}{c} A \mapsto \text{w} \\ B \mapsto \text{r}\end{array}$};
						\node[markedInvalidated] (AwBw) at (5.5,4) {$\begin{array}{c} A \mapsto \text{w} \\ B \mapsto \text{r}\end{array}$};
						\node[markedInvalidated] (AwBrw) at (7,2.5) {$\begin{array}{c} A \mapsto \text{w}\\ B \mapsto \text{rw}\end{array}$};

						\draw[edge] (none) -- (Ar);
						\draw[edge] (none) -- (Aw);
						\draw[edge] (none) -- (Br);
						\draw[edge] (none) -- (Bw);

						\draw[edge] (Ar) -- (Arw);
						\draw[edge] (Aw) -- (Arw);
						\draw[edge] (Br) -- (Brw);
						\draw[redEdge] (Bw) -- (Brw);

						\draw[edge] (Ar) -- (ArBr);
						\draw[edge] (Br) -- (ArBr);
						\draw[edge] (Ar) -- (ArBw);
						\draw[redEdge] (Bw) -- (ArBw);
						\draw[edge] (Ar) -- (ArBrw);
						\draw[redEdge] (Brw) -- (ArBrw);

						\draw[edge] (Aw) -- (AwBr);
						\draw[edge] (Br) -- (AwBr);
						\draw[edge] (Aw) -- (AwBw);
						\draw[redEdge] (Bw) -- (AwBw);
						\draw[edge] (Aw) -- (AwBrw);
						\draw[redEdge] (Brw) -- (AwBrw);

						\draw[redhalfInvalidatedDottedEdge] (Arw) -- (ArwBr);
						\draw[edge] (Br) -- (ArwBr);
						\draw[redhalfInvalidatedDottedEdge] (Arw) -- (ArwBw);
						\draw[redEdge] (Bw) -- (ArwBw);
						\draw[redhalfInvalidatedDottedEdge] (Arw) -- (ArwBrw);
						\draw[redEdge] (Brw) -- (ArwBrw);

				\end{tikzpicture}
			}
			\vspace*{-1.5em}
			\caption{Compromise Graph for \acrshort{arh}-Verification Traversal}
			\label{fig:traversal-algorithm}
			\vspace{-2em}
		\end{wrapfigure}
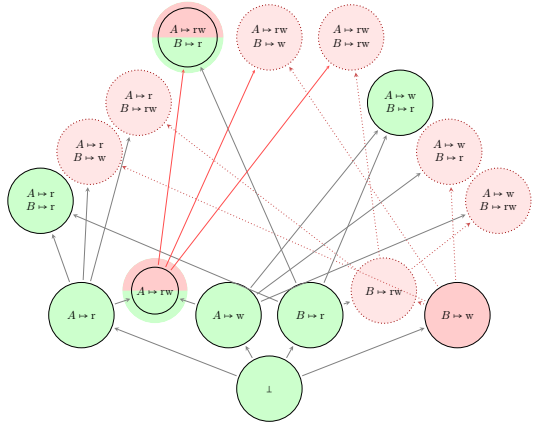
		In our case study, in which we have 4~entities (\glspl{ecu}), 3~domains, 2~permissions, and 2~\gls{sp}, we require the  verification of $(2^2)^{4+3} \cdot 2 = 32,768$ lemmas (i.e. individual \glspl{sp} to be validated against specific compromise scenarios).
		Even under the optimistic (and unrealistic) assumption that each verification takes only one second,\footnote{\Gls{tamarin} process has startup and other overheads.} the total runtime is roughly $32,768\,\text{s} > 9\,\text{h}$.
		Hence, a naive verification strategy is impractical for complex \gls{arh} analyses and an algorithmic improvement is required.

		Below we present an algorithmic orchestration optimization that enables verification across all compromise scenarios and supports fine-grained capabilities and multiple \glspl{sp}.
		This optimization makes \gls{arh} verification for complex adversarial and test scenarios in automotive network architectures feasible and is a necessary prerequisite for \gls{pm} analyses.

		We exploit the monotonicity of \gls{sp} invalidation with respect to compromised entities and domains and their capabilities.
		If a \gls{sp}~$S$ is invalidated by compromise~$c$, then any~$c'$ with~$c \preceq c'$ also invalidates~$S$~\cite{girolSpectralAnalysisNoise2020}.\\
		Instead of verifying all scenarios naively, we can thus simply identify the minimal scenarios~$\text{MCS}$ that cause violations.
		To this end, we perform a preorder traversal of the Hasse diagram (see Fig.~\ref{fig:traversal-algorithm}) of the partial order~$\preceq$ starting at the least element, in which no entity is compromised.
		At each node we invoke \gls{tamarin} to verify the \glspl{sp} (lemmas).
		If all \glspl{sp} hold, then the traversal proceeds.
		Otherwise, for any \gls{sp} invalidated, we simply mark all larger (with respect to~$\preceq$) scenarios as equally invalidated for that property and skip their (costly) verification.
		This optimization leverages monotonicity under the compromise order~$\preceq$ and thus avoids redundant checks.
		Traversal then resumes, and at each node we verify only the properties not already marked invalidated.

	\subsection{Limitations}
		The \gls{fv} of \glspl{ana} and their protocols with \glspl{arh} enables comprehensive, comparative analysis of complex \gls{sp}.
		For complex compromise scenarios, our approach assigns responsibility for \gls{sp}'s invalidation.
		We identify which components, under which conditions (read or write access), influence \gls{sp} invalidation, i.e. \emph{who} is responsible.
		A limitation of the current approach is the \emph{how}: which adversarial interactions cause invalidation in each scenario.
		Our approach yields many traces across all invalidated scenarios; each scenario can produce multiple distinct traces, which are numerous and difficult to analyze manually.

	\section{Comprehensive Adversarial Behavior Analysis}\label{chap:adversarial-behavior-analysis}
	We address the main limitation of \gls{arh}, enabling deducing the \emph{how} in addition to the \emph{who}, by introducing an approach to \gls{pm} of hypertraces.
	Our method summarizes attacker behavior leading to \gls{sp} invalidation, using \gls{pm} as an additional analysis step based on (\gls{arh}) execution traces that invalidate \gls{sp}.\\

	\subsection{Formalization}\label{chap:process-mining-formalization}
		Below, we formalize the components relevant for applying \gls{pm} to execution traces.
		\gls{pm} concepts in this section are based on~\cite{vanderaalstProcessMiningHandbook2022}.
		The analysis starts from \emph{execution traces}~$w \in M^*$ (see Sect.~\ref{chap:identifying-impact-component-compromise}), which we convert and aggregate to an event-log.

		We define the valid sets of execution traces~$L(S)$ that invalidate a specific \gls{sp}~$S$ as~$L(S) = \{\, w \in M^* \mid \exists\, c \,\colon E \to P \colon w \not\models_c S \text{ and } w \text{ is valid for } c \,\}$.
		Prior to aggregating execution traces into an event-log, we formalize the \gls{pm} components relevant for our analysis.
		An \emph{event}~$\epsilon = \langle a, \mathrm{id}, t\rangle$ is an atomic activity occurrence that contains an activity name~$a \in \mathbb{A}$ from a set of possible activities~$\mathbb{A}$, a case identifier~$\mathrm{id} \in \mathbb{N}$ that uniquely identifies the process instance to which the event is assigned, and a time-stamp~$t \in \mathbb{R}$.
		We let~$\id(\epsilon)=\mathrm{id}$ be the identifier of an event~$\epsilon = \langle a, \mathrm{id}, t\rangle$.

		The set of all possible events is denoted by~$\mathcal{E}$; every~$\epsilon \in \mathcal{E}$ is an atomic, time-stamped activity occurrence tied to exactly one case identifier.\\
		An \emph{event trace}~$\sigma$ is a finite sequence~$\sigma = \langle \epsilon_1, \epsilon_2, \dotsc, \epsilon_n \rangle$ of events~$\epsilon_i \in \mathcal{E}$ for all~$1 \leq i \leq n$ such that all events in~$\sigma$ belong to the same case; i.e., $\id(\epsilon_i) = \id(\epsilon_j)$ for all $1 \leq i,j \leq n$.
		Thus we also write~$\id(\sigma) = \id(\epsilon_1)$ for the case of the trace.\\
		An event trace denotes a single process instance and the set of all possible traces is denoted by~$\mathcal{E}^*$.\\
		An \emph{event-log}~$\mathcal{L}$ is a set $\mathcal{L} = \{\sigma_1,\sigma_2,\dots,\sigma_n\}$ of event traces.
		Different traces in an event-log represent unique cases: for any~$\sigma, \sigma' \in \mathcal{L}$ with~$\sigma \neq \sigma'$ we have $\id(\sigma) \neq \id(\sigma')$.

		A \emph{\acrlong{pm} algorithm}~$\mathcal{MA}$ maps event-logs to process models: \linebreak$\mathcal{MA} \colon \mathcal{P}(\mathcal{E}^*) \to \mathcal{M}$, where~$\mathcal{P}(\mathcal{E}^*)$ is the set of all event-logs (sets of event traces) and~$\mathcal{M}$ is the set of all process models.
		We abstract internal details and treat~$\mathcal{MA}$ as a black box that returns a model~$m \in \mathcal{M}$ for a given log~$\mathcal{L}$.\\
		We define the function~$f_a \colon M \to \mathbb{A}$ to map every message~$m$ in an execution trace~$w\in M^*$ to an activity name~$f_a(m) \in\mathbb{A}$.
		For modeling convenience, we simply let the \emph{activity label} be the concatenation of the message components, so for a message~$m=((s,r),c)$ we set~$f_a(m) =s \circ r\circ c$.
		Let~$f_{\mathrm{id}} \colon M^* \to \mathbb{N}$ return a unique non-negative integer (identifier) for any execution trace~$w \in M^*$.
		Finally, let $\Delta_t > 0$ be some arbitrary increment.\\
		We define the \emph{transformation}~$\mathcal{T} \colon M^* \to \mathcal{E}^*$ to map an execution trace \linebreak$w=\langle m_1, \dotsc, m_n\rangle$ and its unique case identifier~$i = f_{\mathrm{id}}(w)$ to an event trace
		\[ \mathcal{T}(w) = \Bigl\langle \langle f_a(m_1), i, 1 \cdot \Delta_t\rangle, \dotsc, \langle f_a(m_n),i, n \cdot \Delta_t\rangle \Bigr\rangle. \]
		Let~$L(S)= \{w_1, \dotsc, w_n\}$ be the finite set of distinct execution traces obtained by formal verification that invalidate the \gls{sp}~$S \in \mathtt{S}$.
		Each~$w \in L(S)$ is a distinct counterexample to~$S$.
		To create a \emph{synthetic event-log}~$\mathcal{SL}(S)$, we transform the execution traces of $L(S)$ via~$\mathcal{T}$ to an event-log.
		The synthetic event-log~$\mathcal{SL}(S)$ is the following set of event traces.
		\[
			\mathcal{SL}(S) = \{\, \mathcal{T}(w) \mid w \in L(S) \,\}
		\]
		This event-log $\mathcal{SL}(S)$ is the input to the \gls{pm} algorithm~$\mathcal{MA}$ and yields a process model~$\mathcal{M} = \mathcal{MA}(\mathcal{SL}(S))$.
		\[
			L(S)\ \xrightarrow{\ \mathcal{T}\ }\ \mathcal{SL}(S)
			\ \xrightarrow{\ \mathcal{MA}\ }\ \mathcal{M}\,.
		\]
		In summary, our method bridges \emph{formal verification} and \emph{\acrlong{pm}} via a dedicated pipeline: counterexample traces~$L(S)$ are aggregated into a synthetic event-log~$\mathcal{SL}(S)$ using the transformation function~$\mathcal{T}$, and a \gls{pm} algorithm~$\mathcal{MA}$ maps~$\mathcal{SL}(S)$ to a process model~$\mathcal{M}$.

	\subsection{Classification and Limitations of the Approach}
		Our approach provides an exploratory contribution to applying \gls{pm} for security-focused insights from \gls{fv} results.
		To our knowledge, there is no previous related work in this area (see Sect. \ref{chap:related-work}).\\
		We leverage \gls{pm} techniques to achieve high replay fitness and precision, so the discovered model reflects the event-logs and permits only observed behavior (or behavior representing filtered simplifications).
		Model simplicity should be balanced against the complexity needed to capture real adversarial behavior, which makes the generalization less critical.\\
		Furthermore, our overarching goal diverges from typical \gls{pm} objectives.
		Our focus is on identifying behavior patterns, individual or collective, rather than scrutinizing entire process flows.
		Thus, depicting a complete, flawless process flow is out of scope.
		Understanding relevant (attacker) behavior suffices to address and mitigate potential invalidation.

	\section{Prototypical Implementation}\label{chap:prototype}
\begin{wrapfigure}{r}{0.45\textwidth}
			\vspace{-0em}
			\centering
			\resizebox{0.45\textwidth}{!}{
				\hspace*{-4em}
				\tikzstyle{model} = [rectangle, rounded corners, minimum width=3cm, minimum height=1cm,text centered, draw=black, fill=red!30]
				\tikzstyle{process} = [rectangle, minimum width=3cm, minimum height=1cm, text centered, draw=black, fill=orange!30]
				\tikzstyle{verification} = [rectangle, minimum width=3cm, minimum height=1cm, text centered, draw=black, fill=orange!30]
				\tikzstyle{enriched} = [rectangle, rounded corners, minimum width=3cm, minimum height=1cm, text centered, draw=black, fill=red!30!purple!30]
				\tikzstyle{results} = [rectangle, minimum width=3cm, minimum height=1cm, text centered, draw=black, fill=green!30!]
				\tikzstyle{tamarin} = [rectangle, minimum width=3cm, minimum height=1cm, text centered, draw=black, fill=green!30]
				\tikzstyle{postprocessing} = [rectangle, minimum width=3cm, minimum height=1cm, text centered, draw=black, fill=orange!30]
				\tikzstyle{processmining} = [rectangle, minimum width=3cm, minimum height=1cm, text centered, draw=black, fill=cyan!30]
				\tikzstyle{eventlog} = [rectangle, minimum width=3cm, minimum height=1cm, text centered, draw=black, fill=blue!20]
				\tikzstyle{arrow} = [thick, -{Stealth[scale=1.2]}]
				\begin{tikzpicture}[node distance=2cm]
						\tikzstyle{every node}=[font=\large]

						\node (model) [model] {
							\begin{minipage}{2.8cm}
								\centering
								Architecture \&\\ Protocol Model\\\vspace*{0.2cm}
								\includegraphics[width=1.8cm]{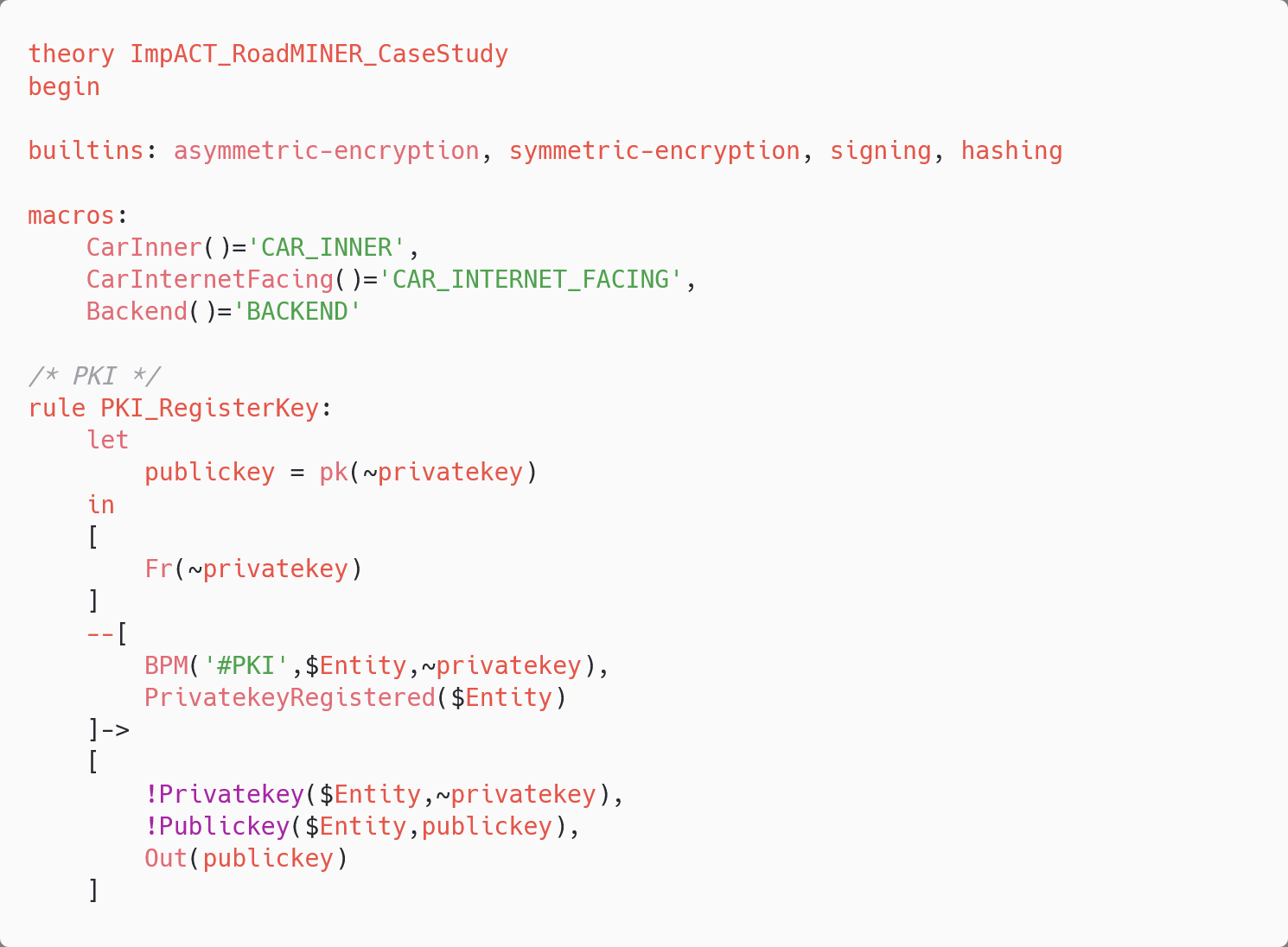}\\
							\end{minipage}
						};

						\node (algorithmic) [process, below of=model, xshift=0cm, yshift=-1cm, minimum width=2cm, minimum height=1.5cm] {
							\begin{minipage}{2.5cm}
								\centering
								Algorithmic Verification\\\vspace*{0.2cm}
								\includegraphics[width=2cm]{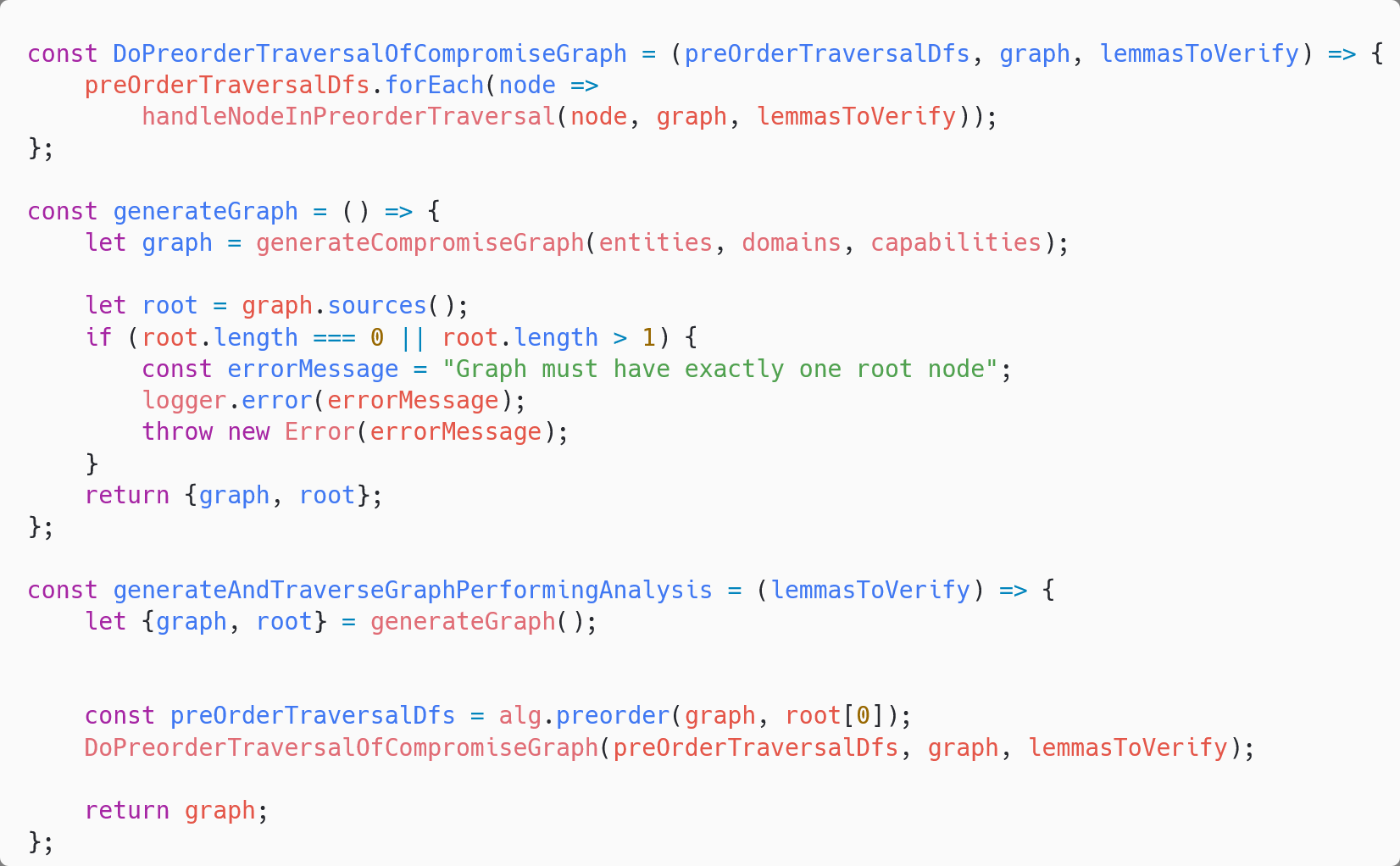}
							\end{minipage}
						};
						\draw [arrow] (model) -- (algorithmic);

						\node (preprocessing3) [process, below of=algorithmic, xshift=0.15cm, yshift=-1.1cm, minimum width=2cm, minimum height=1.5cm] {
							\begin{minipage}{2.5cm}
								\centering
								Preprocessing\\\vspace*{0.2cm}
								\includegraphics[width=2cm]{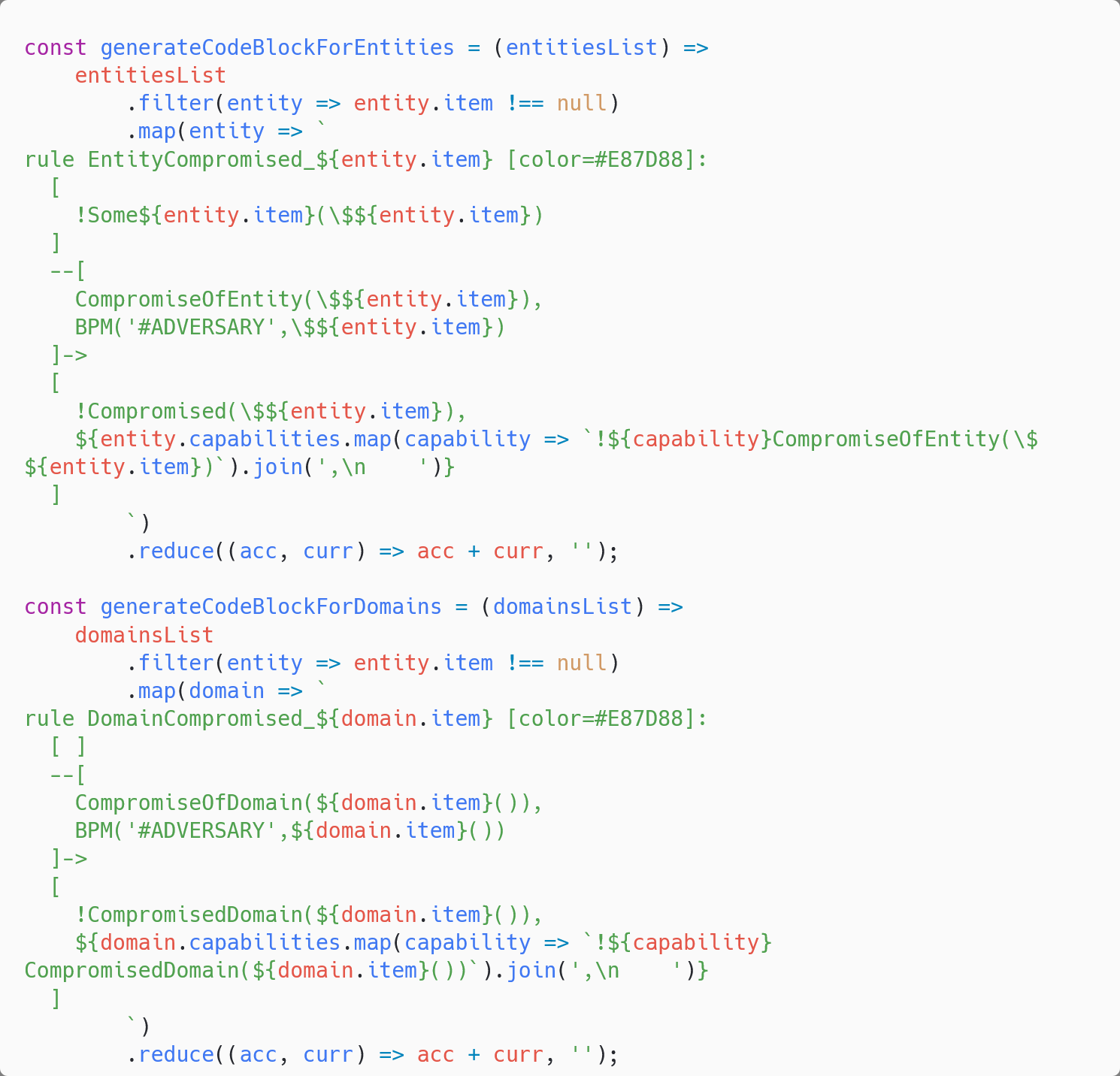}
							\end{minipage}
						};
						\draw [arrow] (algorithmic) -- (preprocessing3);

						\node (preprocessing2) [process, below of=algorithmic, xshift=0cm, yshift=-1cm, minimum width=2cm, minimum height=1.5cm] {
							\begin{minipage}{2.5cm}
								\centering
								Preprocessing\\\vspace*{0.2cm}
								\includegraphics[width=2cm]{./res/preprocessing-invert}
							\end{minipage}
						};
						\draw [arrow] (algorithmic) -- (preprocessing2);

						\node (preprocessing1) [process, below of=algorithmic, xshift=-0.15cm, yshift=-0.9cm, minimum width=2cm, minimum height=1.5cm] {
							\begin{minipage}{2.5cm}
								\centering
								Preprocessing\\\vspace*{0.2cm}
								\includegraphics[width=2cm]{./res/preprocessing-invert}
							\end{minipage}
						};
						\draw [arrow] (algorithmic) -- (preprocessing1);

						\node (enrichedmodel3) [enriched, below of=preprocessing3, xshift=0cm, yshift=-1.2cm, minimum width=2.5cm, minimum height=1.5cm] {
							\begin{minipage}{2.5cm}
								\centering
								Enriched Model\\\vspace*{0.2cm}
								\includegraphics[width=1.8cm]{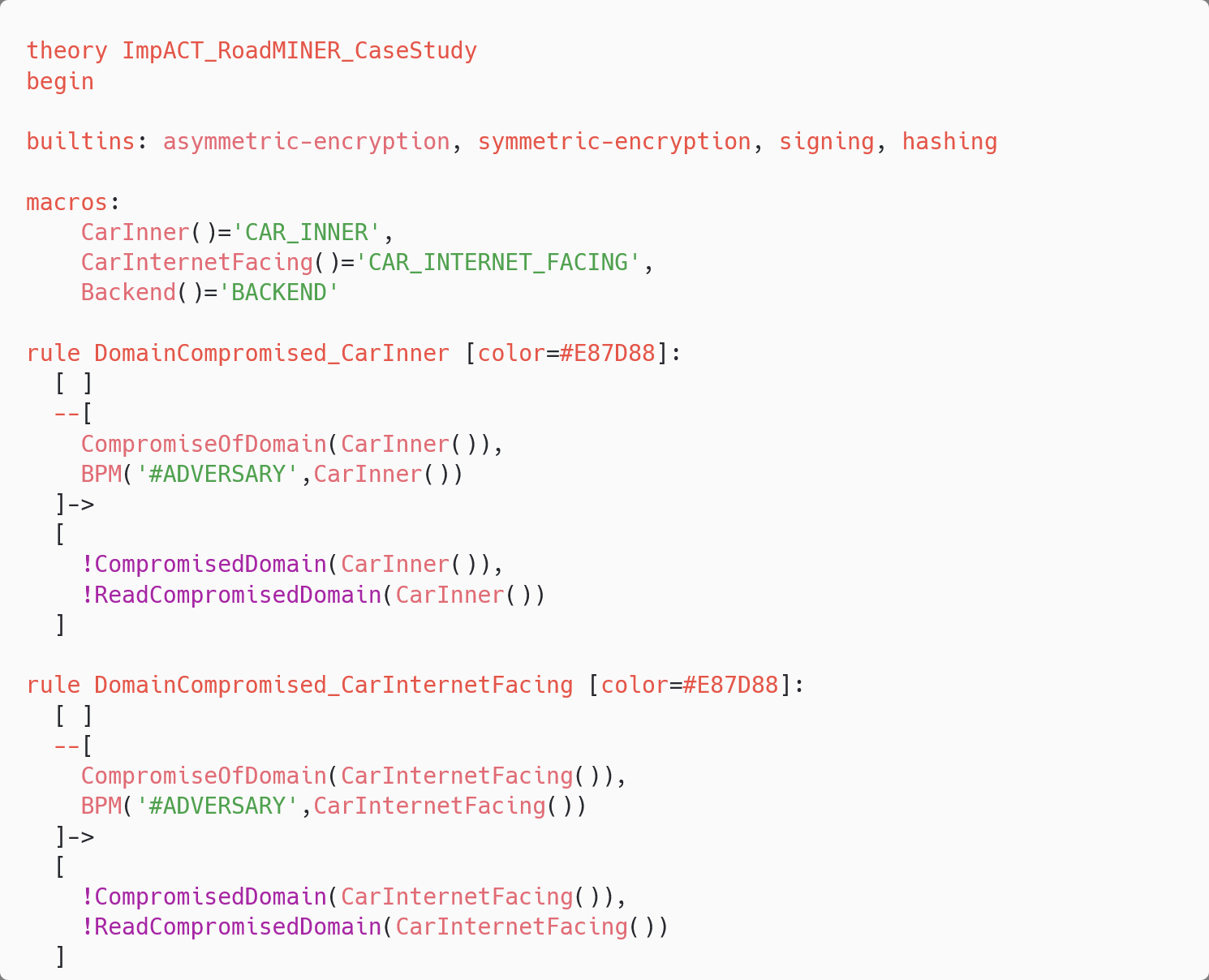}
							\end{minipage}
						};
						\draw [arrow] (preprocessing3) -- (enrichedmodel3);
						\node (enrichedmodel2) [enriched, below of=preprocessing2, xshift=0cm, yshift=-1.2cm, minimum width=2.5cm, minimum height=1.5cm] {
							\begin{minipage}{2.5cm}
								\centering
								Enriched Model\\\vspace*{0.2cm}
								\includegraphics[width=1.8cm]{./res/theory-enriched-inverted}
							\end{minipage}
						};
						\draw [arrow] (preprocessing2) -- (enrichedmodel2);
						\node (enrichedmodel1) [enriched, below of=preprocessing1, xshift=0cm, yshift=-1.2cm, minimum width=2.5cm, minimum height=1.5cm] {
							\begin{minipage}{2.5cm}
								\centering
								Enriched Model\\\vspace*{0.2cm}
								\includegraphics[width=1.8cm]{./res/theory-enriched-inverted}
							\end{minipage}
						};
						\draw [arrow] (preprocessing1) -- (enrichedmodel1);

						\node (verification3) [verification, below of=enrichedmodel3, xshift=0cm, yshift=-1cm, minimum width=2cm, minimum height=1.5cm] {
							\begin{minipage}{2.5cm}
								\centering
								Controlled Verification\\\vspace*{0.2cm}
								\includegraphics[width=2cm]{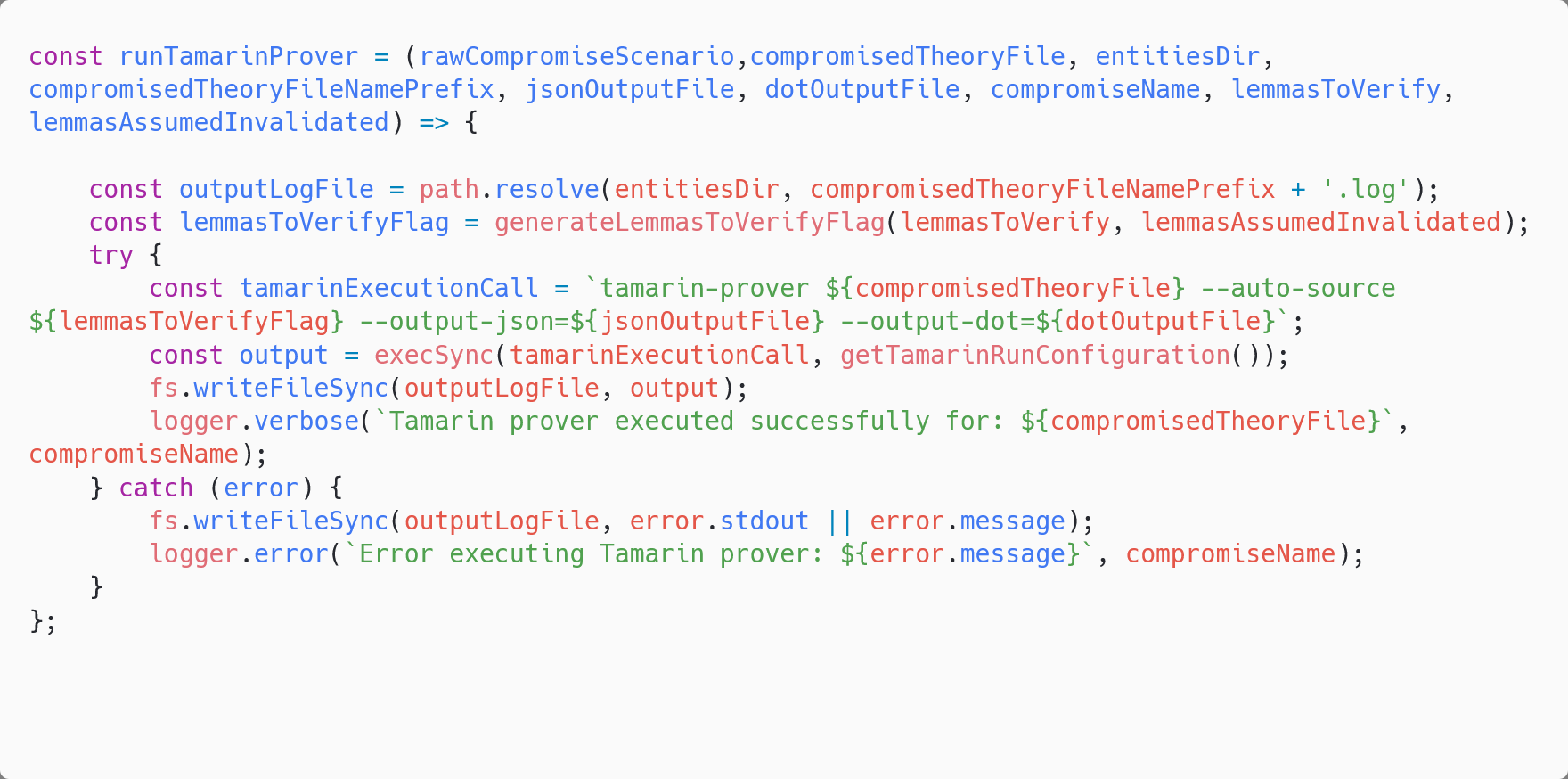}
							\end{minipage}
						};
						\draw [arrow] (enrichedmodel3) -- (verification3);
						\node (verification2) [verification, below of=enrichedmodel2, xshift=0cm, yshift=-1cm, minimum width=2cm, minimum height=1.5cm] {
							\begin{minipage}{2.5cm}
								\centering
								Controlled Verification\\\vspace*{0.2cm}
								\includegraphics[width=2cm]{./res/controlled-verification-inverted}
							\end{minipage}
						};
						\draw [arrow] (enrichedmodel2) -- (verification2);
						\node (verification1) [verification, below of=enrichedmodel1, xshift=0cm, yshift=-1cm, minimum width=2cm, minimum height=1.5cm] {
							\begin{minipage}{2.5cm}
								\centering
								Controlled Verification\\\vspace*{0.2cm}
								\includegraphics[width=2cm]{./res/controlled-verification-inverted}
							\end{minipage}
						};
						\draw [arrow] (enrichedmodel1) -- (verification1);

						\node (tamarin) [tamarin, right of=verification2, xshift=1.75cm, yshift=0cm, minimum width=2cm, minimum height=1.5cm] {
							\begin{minipage}{3cm}
								\centering
								Tamarin Prover\\\vspace*{0.2cm}
								\includegraphics[width=1.5cm]{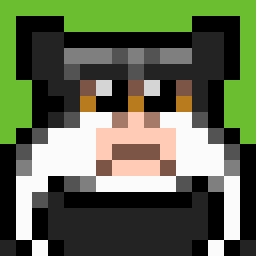}
							\end{minipage}
						};
						\draw [arrow] (verification3) -- (tamarin);
						\draw [arrow] (verification2) -- (tamarin);
						\draw [arrow] (verification1) -- (tamarin);

						\node (output3) [results, above of=tamarin, xshift=0.15cm, yshift=0.9cm, minimum width=3cm, minimum height=1.5cm] {
							\begin{minipage}{2.5cm}
								\centering
								Verification Results\\\vspace*{0.15cm}
								\includegraphics[width=2cm]{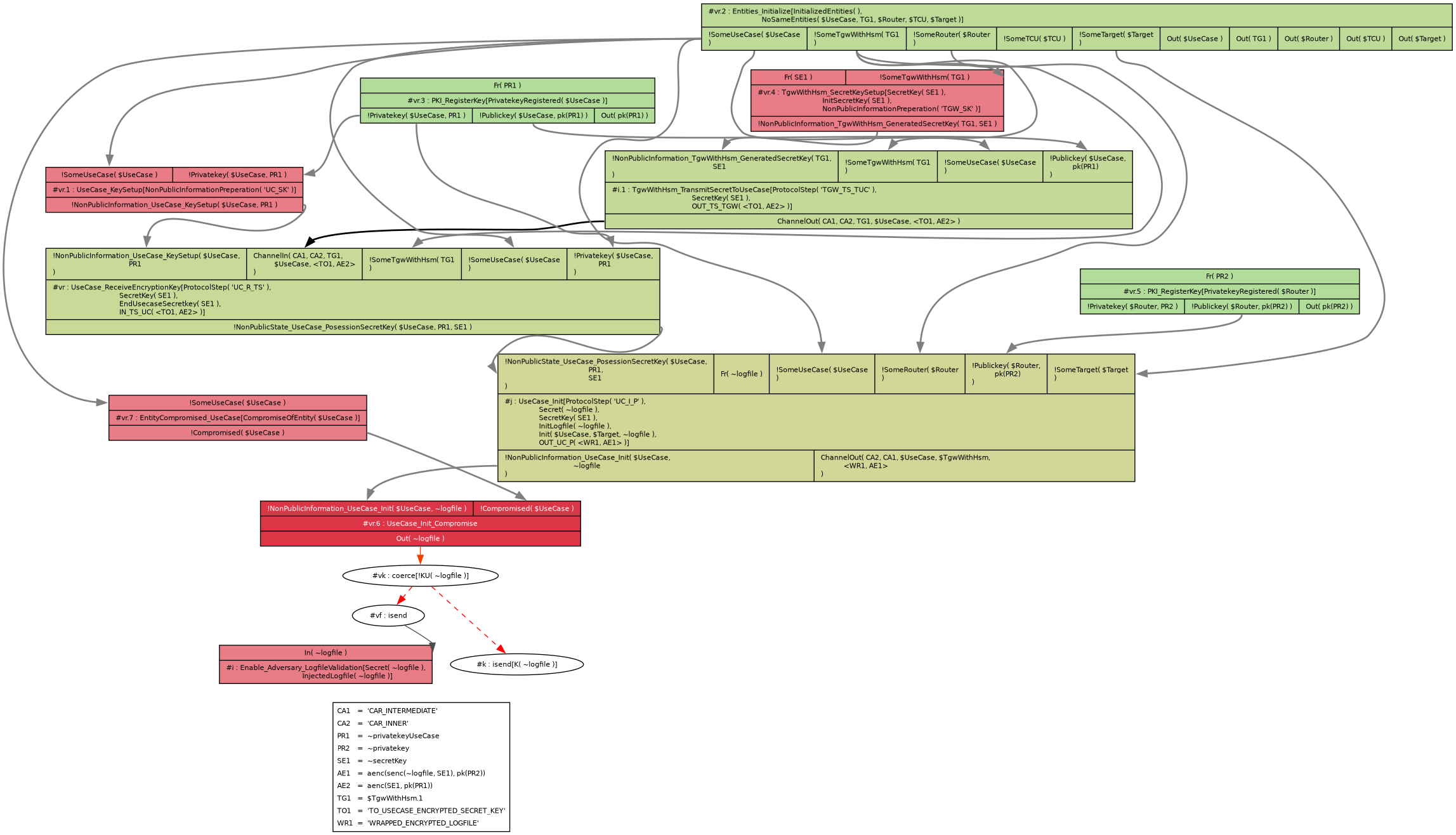}
							\end{minipage}
						};
						\node (postprocessing3) [postprocessing, above of=output3, xshift=0cm, yshift=1cm, minimum width=3cm, minimum height=1.5cm] {
							\begin{minipage}{3cm}
								\centering
								Postprocessing\\\vspace*{0.2cm}
								\includegraphics[width=2.5cm]{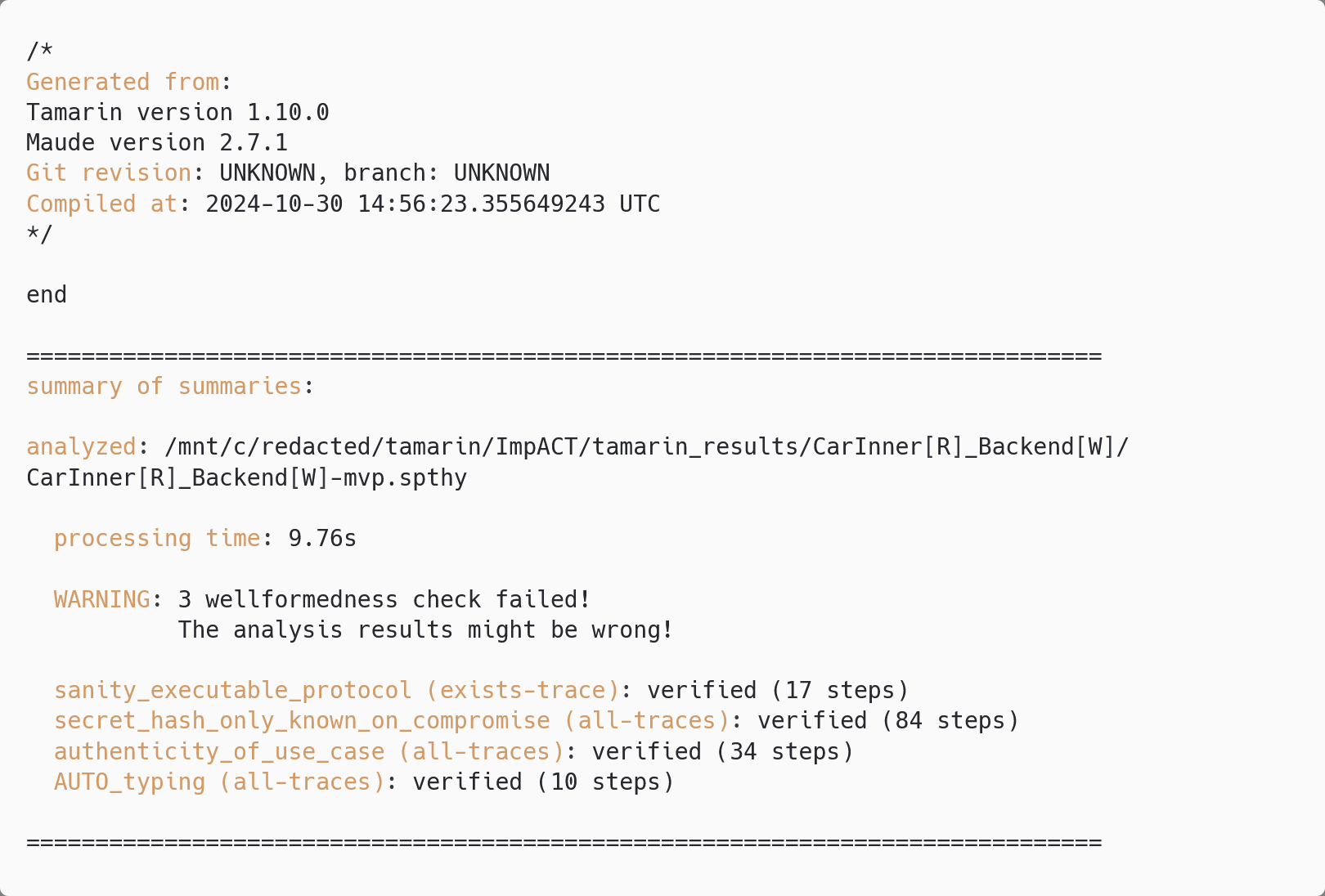}
							\end{minipage}
						};
						\draw [arrow] (output3) -- (postprocessing3);

						\node (output2) [results, above of=tamarin, xshift=0cm, yshift=1cm, minimum width=3cm, minimum height=1.5cm] {
							\begin{minipage}{2.5cm}
								\centering
								Verification Results\\\vspace*{0.15cm}
								\includegraphics[width=2cm]{./res/invalidation}
							\end{minipage}
						};
						\node (postprocessing2) [postprocessing, above of=output2, xshift=0cm, yshift=1cm, minimum width=3cm, minimum height=1.5cm] {
							\begin{minipage}{3cm}
								\centering
								Postprocessing\\\vspace*{0.2cm}
								\includegraphics[width=2.5cm]{./res/results-inverted}
							\end{minipage}
						};
						\draw [arrow] (output2) -- (postprocessing2);

						\node (output1) [results, above of=tamarin, xshift=-0.15cm, yshift=1.1cm, minimum width=3cm, minimum height=1.5cm] {
							\begin{minipage}{2.5cm}
								\centering
								Verification Results\\\vspace*{0.15cm}
								\includegraphics[width=2cm]{./res/invalidation}
							\end{minipage}
						};
						\node (postprocessing1) [postprocessing, above of=output1, xshift=0cm, yshift=1cm, minimum width=3cm, minimum height=1.5cm] {
							\begin{minipage}{3cm}
								\centering
								Postprocessing\\\vspace*{0.2cm}
								\includegraphics[width=2.5cm]{./res/results-inverted}
							\end{minipage}
						};
						\draw [arrow] (output1) -- (postprocessing1);

						\node (arh-extraction-score) [process, above of=postprocessing2, xshift=0cm, yshift=1.2cm, minimum width=2cm, minimum height=1.5cm] {
							\begin{minipage}{3cm}
								\centering
								ARH-Extraction\\\vspace*{0.2cm}
								\includegraphics[height=1.5cm]{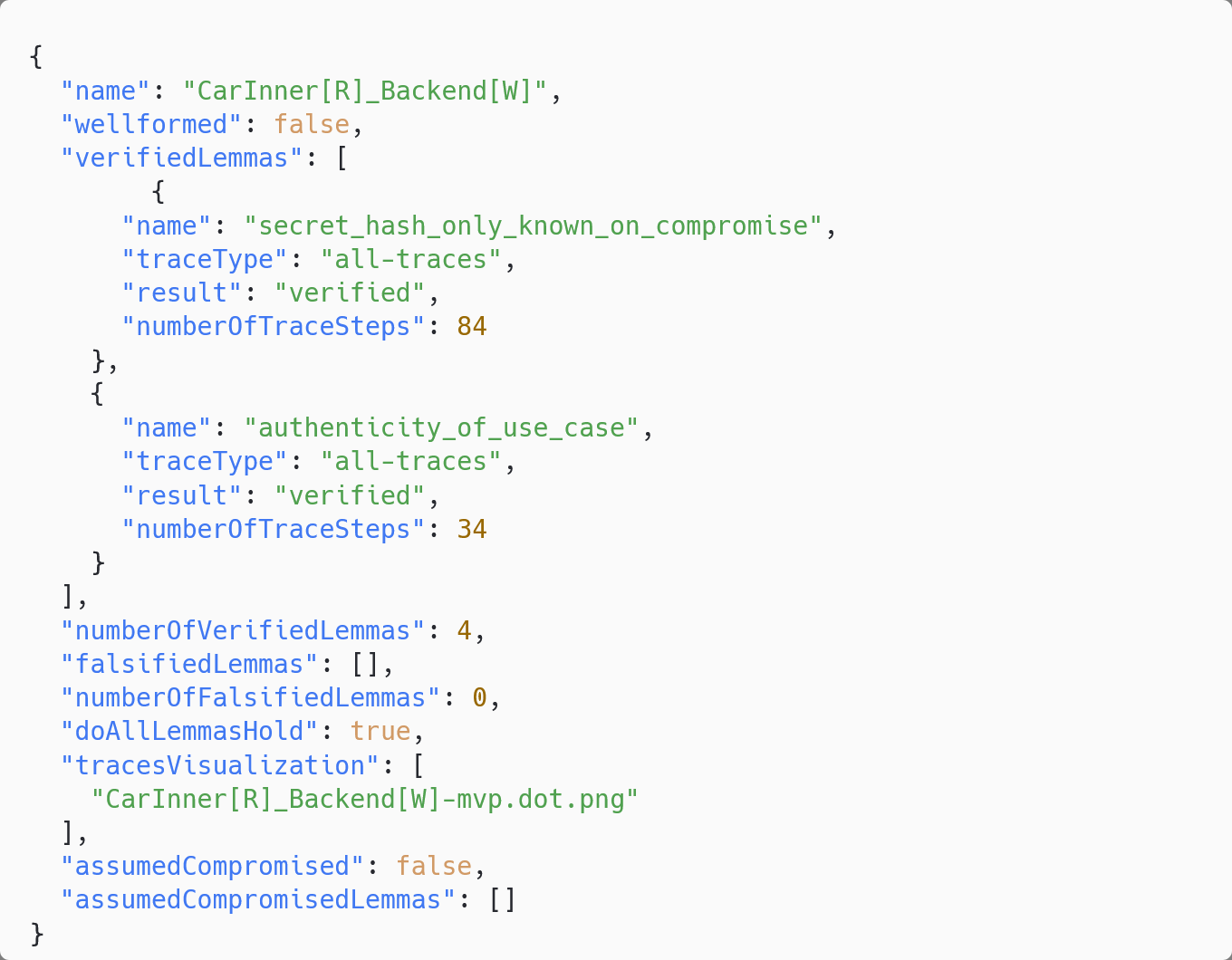}
							\end{minipage}
						};

						\draw [arrow] (postprocessing3) -- (arh-extraction-score);
						\draw [arrow] (postprocessing2) -- (arh-extraction-score);
						\draw [arrow] (postprocessing1) -- (arh-extraction-score);

						\draw [arrow] (postprocessing3) to[out=165, in=-20] (algorithmic);
						\draw [arrow] (postprocessing2) to[out=165, in=-20] (algorithmic);
						\draw [arrow] (postprocessing1) to[out=165, in=-20] (algorithmic);

						\node (output) [process, above of=arh-extraction-score, xshift=0cm, yshift=1cm, minimum width=2cm, minimum height=1.5cm] {
							\begin{minipage}{3cm}
								\centering
								Output and Visualization\\\vspace*{0.2cm}
								\includegraphics[width=2.5cm]{./res/ImpACT-GUI_highDpi}
							\end{minipage}
						};

						\draw[dashed, thick, draw=orange] ($(output.north east)+(0.2,0.2)$) -- ($(output.north west)+(-0.2,0.2)$);
						\draw[dashed, thick, draw=orange] ($(output.north west)+(-0.2,0.2)$) -- ($(algorithmic.north east)+(0.55,0.2)$);
						\draw[dashed, thick, draw=orange] ($(algorithmic.north west)+(-0.3,0.2)$) -- ($(algorithmic.north east)+(0.55,0.2)$);
						\draw[dashed, thick, draw=orange] ($(algorithmic.north west)+(-0.3,0.2)$) -- ($(verification2.south west)+(-0.4,-0.5)$);
						\draw[dashed, thick, draw=orange] ($(verification2.south east)+(0.4,-0.4)$) -- ($(verification2.south west)+(-0.4,-0.4)$);
						\draw[dashed, thick, draw=orange] ($(verification2.south east)+(0.4,-0.4)$) -- ($(postprocessing2.south west)+(-0.3,-0.3)$);
						\draw[dashed, thick, draw=orange] ($(postprocessing2.south east)+(0.3,-0.3)$) -- ($(postprocessing2.south west)+(-0.3,-0.3)$);
						\draw[dashed, thick, draw=orange] ($(output.north east)+(0.2,0.2)$) -- ($(postprocessing2.south east)+(0.3,-0.3)$);

						\node at ($(output.north) + (0,0.45)$) {\textcolor{orange}{ImpACT}};

						\node (conversion) [processmining, right of=output2, xshift=2cm, yshift=0cm, minimum width=2.5cm, minimum height=1.5cm] {
							\begin{minipage}{3cm}
								\centering
								Event-Log\\Conversion\vspace*{0.2cm}
								\includegraphics[width=1.25cm]{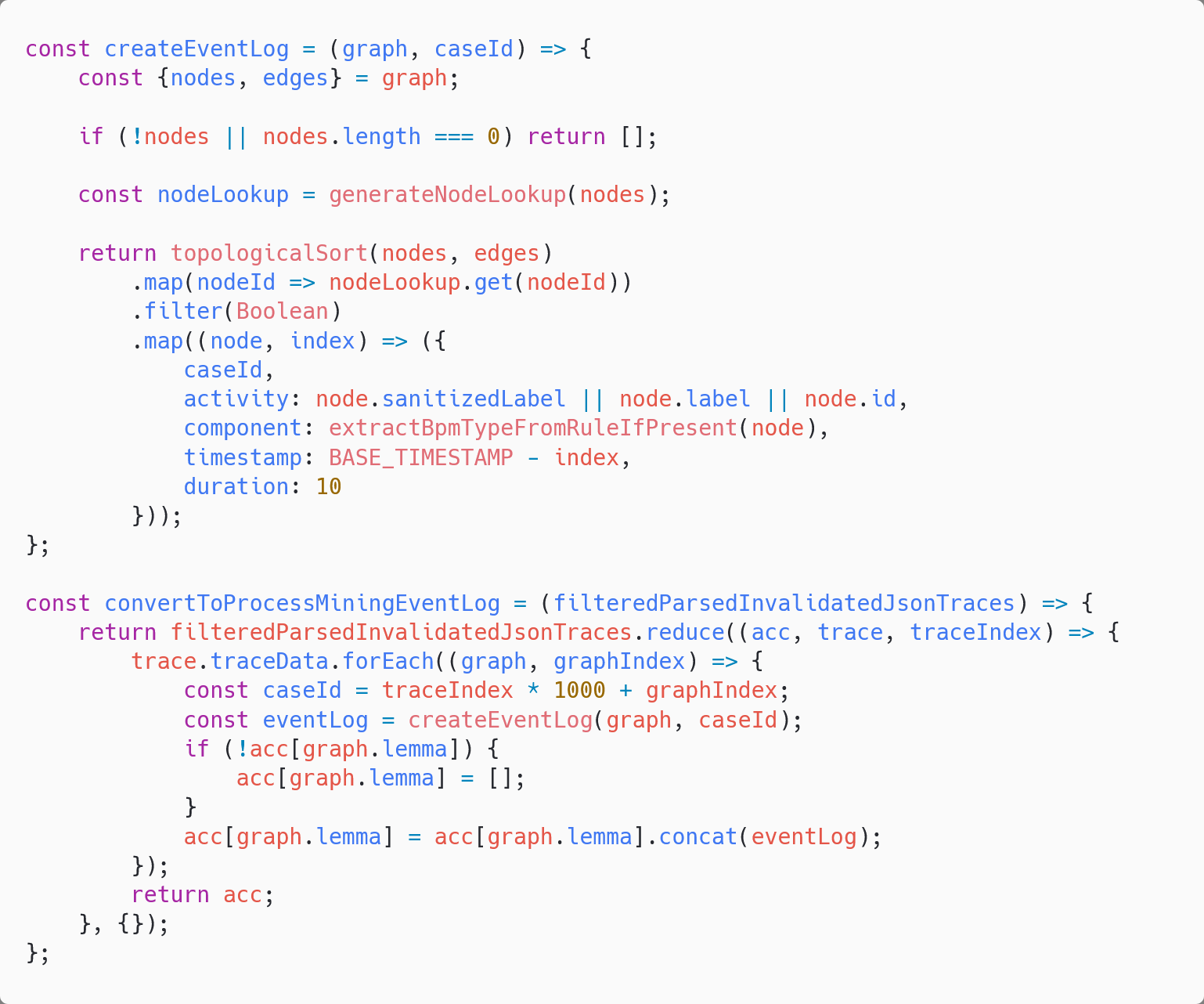}
							\end{minipage}
						};
						\draw [arrow] (postprocessing3) to[out=-30, in=180] (conversion);
						\draw [arrow] (postprocessing2) to[out=-30, in=180] (conversion);
						\draw [arrow] (postprocessing1) to[out=-30, in=180] (conversion);

						\node (eventlogConversion) [eventlog, above of=conversion, xshift=0cm, yshift=1cm, minimum width=2.5cm, minimum height=1.5cm] {
							\begin{minipage}{3cm}
								\centering
								Generated Event-Log\vspace*{0.2cm}
								\includegraphics[width=2cm]{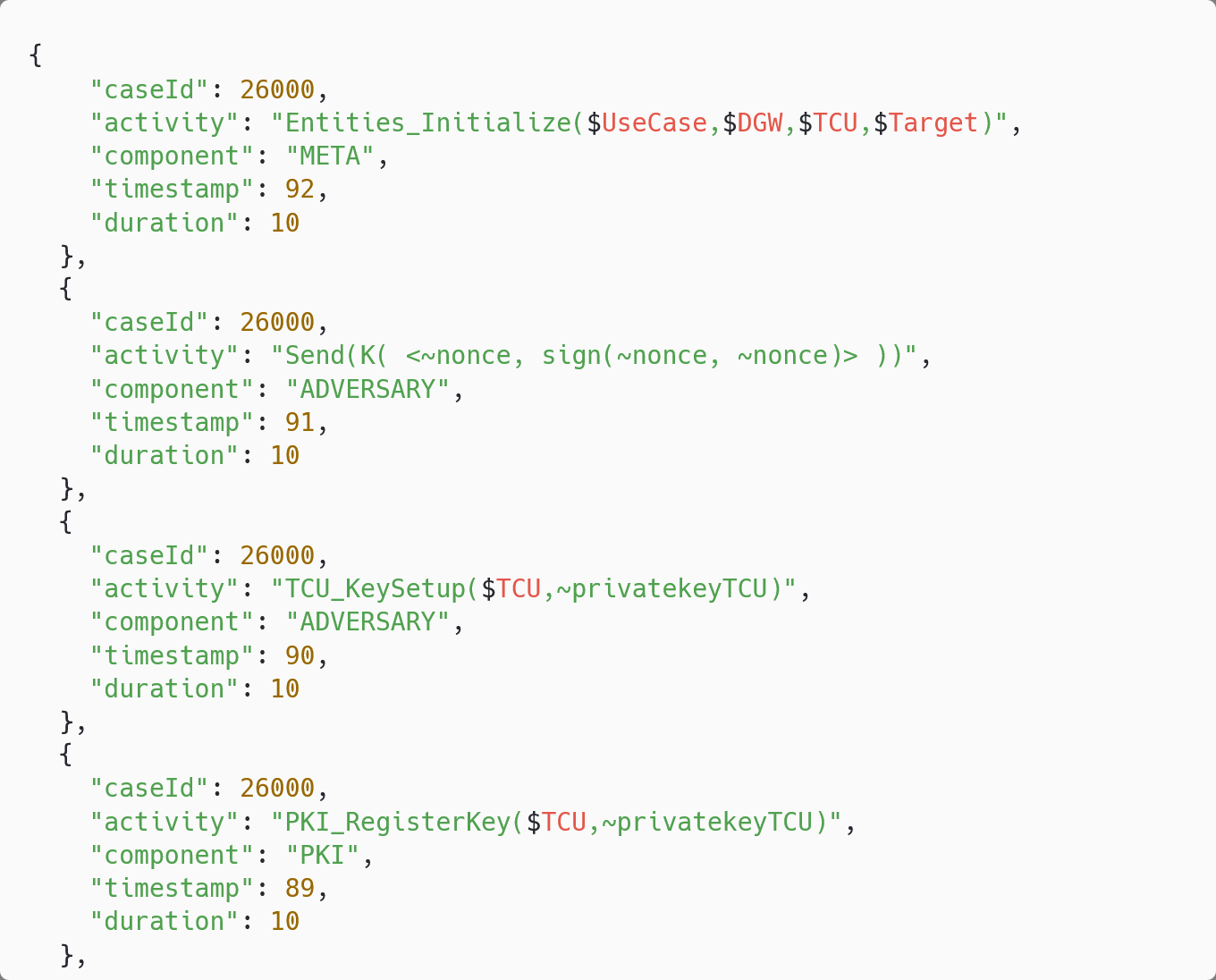}
							\end{minipage}
						};
						\draw [arrow] (conversion) -- (eventlogConversion);

						\node (processMining) [processmining, above of=eventlogConversion, xshift=0cm, yshift=1.1cm, minimum width=2.5cm, minimum height=1.5cm] {
							\begin{minipage}{3cm}
								\centering
								Process Mining\vspace*{0.2cm}
								\includegraphics[width=2.5cm]{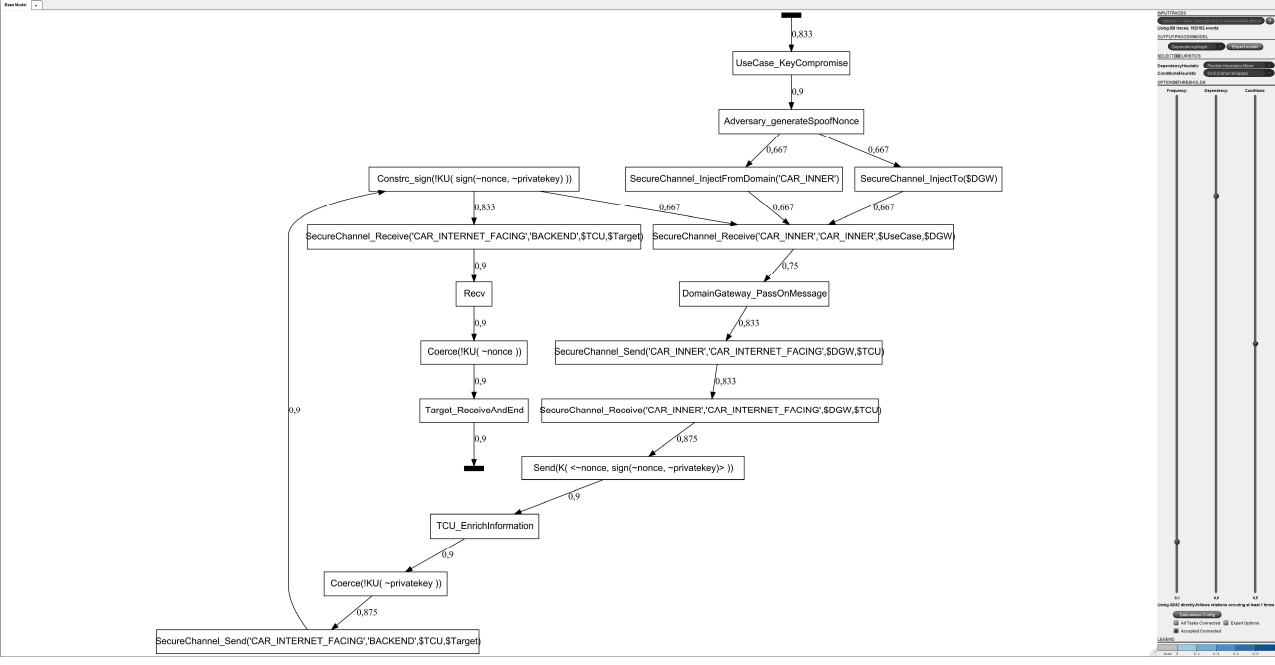}
								\vspace{-4.5em}
							\end{minipage}
						};

						\node (processMiningResults) [processmining, above of=processMining, xshift=0cm, yshift=1cm, minimum width=2.5cm, minimum height=1.5cm] {
							\begin{minipage}{3cm}
								\centering
								Output and\\Visualization\vspace*{0.2cm}
								\colorbox{white}{\includegraphics[width=2.5cm]{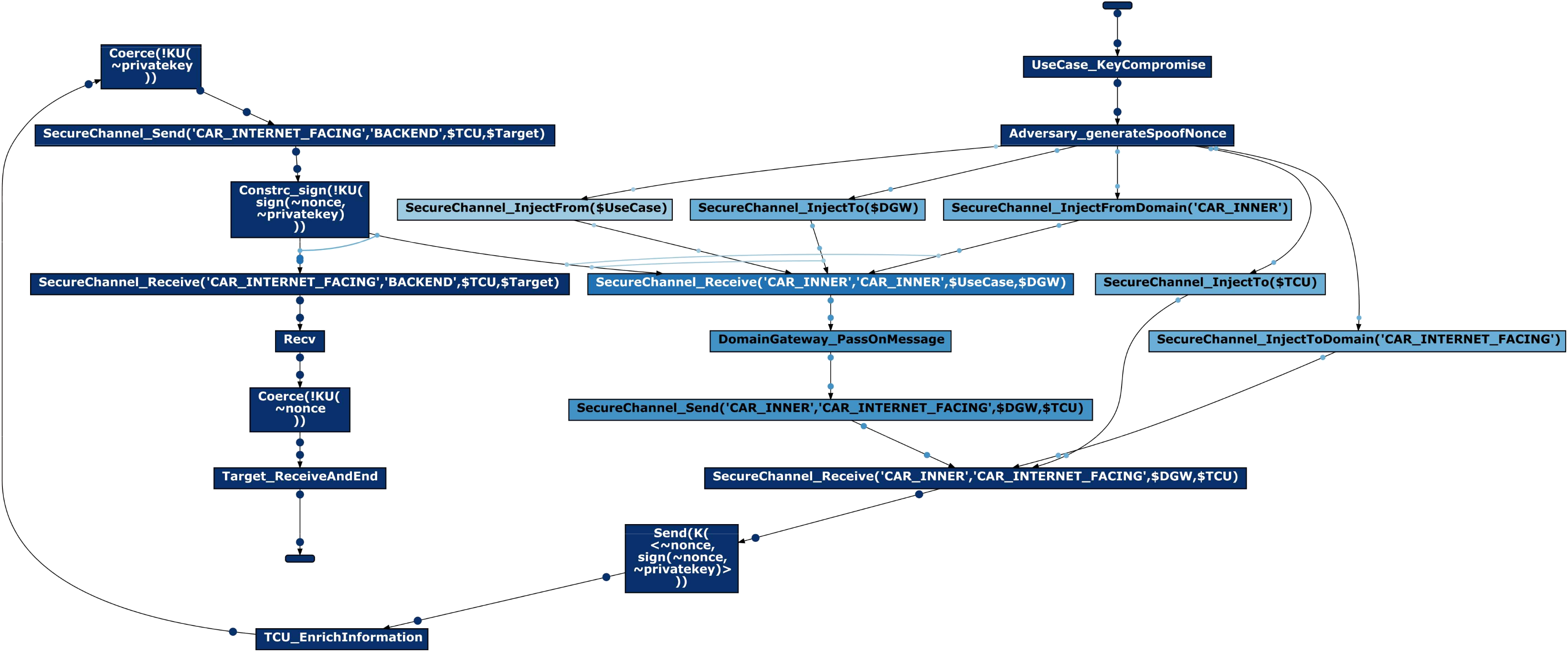}}
							\end{minipage}
						};

						\draw [arrow] (eventlogConversion) -- (processMining);
						\draw [arrow] (processMining) -- (processMiningResults);

						\draw[dashed, thick, draw=cyan] ($(processMiningResults.north west)+(-0.2,0.35)$) rectangle ($(conversion.south east)+(+0.2,-0.2)$);
						\node at ($(processMiningResults.north) + (0,0.65)$) {\textcolor{cyan}{RoadMINER}};

						\draw [arrow] (tamarin) -- (output3);
						\draw [arrow] (tamarin) -- (output2);
						\draw [arrow] (tamarin) -- (output1);

						\draw [arrow] (arh-extraction-score) -- (output);

				\end{tikzpicture}
			}
			\vspace{-1.5em}
			\caption{ImpACT / RoadMINER}
			\label{fig:impact-roadminer-setup}
			\vspace*{-2.5em}
		\end{wrapfigure}

	Our prototypical implementation comprises two components (see Fig. \ref{fig:impact-roadminer-setup}).
	The first is \textit{\acrfull{impact}}, our contribution for in-depth \gls{arh} verification with \gls{tamarin}.
	The second is \textit{\acrfull{roadminer}}, our contribution enabling \gls{pm} on \gls{fv} (hyper)traces.
	We demonstrate their practical use and the resulting substantial benefits through our case study.

	\subsection{\enquote{ImpACT}}
		\begin{wrapfigure}{l}{0.675\textwidth}
			\centering
			\vspace*{-2.25em}
			\includegraphics[width=1\linewidth]{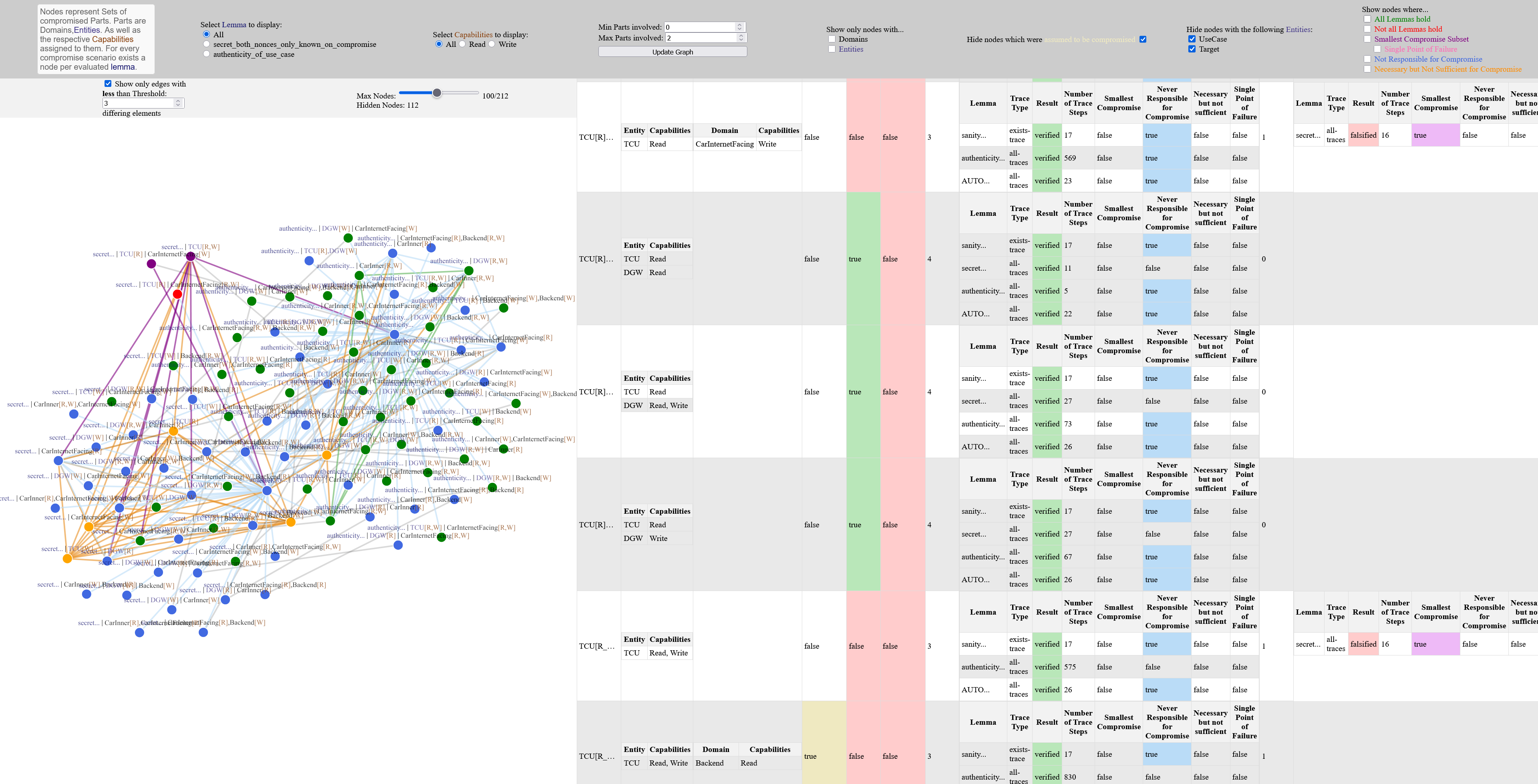}
			\vspace{-1.5em}
			\caption{ImpACT}
			\vspace{-2.75em}
			\label{fig:impact}
		\end{wrapfigure}
		We present our prototype \emph{\acrfull{impact}}, a direct evolution of our previous work \emph{\acrfull{exact}}~\cite{figgeattackresiliencehyperproperties2026}.
		\acrshort{impact}, built on a \emph{Node.js} tech stack, enables fully automated \gls{arh} verification and further analysis by processing interim outputs.
		As a wrapper around \gls{tamarin}, \gls{impact} provides pre- and post-processing for formal models of \glspl{ana} and their protocols, including the \gls{sp} to verify, using \gls{tamarin}'s\emph{ security protocol theory} format as input.

		Execution of \gls{impact} begins with our verification orchestration algorithm (see Sect. \ref{chap:arh-verification-orchetration-algorithm}), generating a graph of all compromise, capability combinations and traversing it in preorder.
		For each scenario (a compromise plus capabilities), pre-processing enriches the base theory with the scenario; controlled verification is then executed, and the results are post-processed via parsing and evaluation.
		A feedback loop uses per-(security)property post-processing results to guide the traversal.
		Consequently, we verify only those \glspl{sp} whose lemmas have not already been marked invalid for a previously examined subset, improving runtime.
		After \gls{tamarin} verification, post-processing converts \gls{tamarin}-specific output into a reusable, machine-readable format augmenting it with analyses and compromise details.
		Using this data, \gls{arh} verification is performed, assessing all \gls{sp} across all scenarios.

		Within the web interface (see Fig. \ref{fig:impact}), \gls{arh} results are presented in a concise overview.
		The interface has two views:
		On the left, is a graph, where each node represents one \gls{arh} result for a given scenario (compromised entities, domains with capabilities) and lemma, and is colored accordingly.
		Nodes with similar characteristics and identical \gls{sp} are clustered via a physics-based\footnote{I.e. gravity simulation via \emph{d3js}} layout, and edges encode order relations between scenarios.
		On the right, a detailed table lists the \gls{arh} results per scenario, including sub-tables for evaluated \glspl{sp} and their associated traces.
		The header bar enables filtering, e.g. by compromise size, included entities/domains, \glspl{sp}, and \gls{arh} outcomes.

	\subsection{\acrshort{roadminer}}
		\emph{\Gls{roadminer}} follows \gls{impact}'s post-processing of \gls{tamarin} results (see Fig.~\ref{fig:impact-roadminer-setup}) to generate synthetic event-logs that capture attacker behavior invalidating \gls{sp}.
		Beyond attribution of compromise focused \gls{arh} analysis, our focus here is the sequence of steps required to violate a given \gls{sp}.\\
		We use verification results, i.e. \gls{sp}-invalidating traces, for event-log creation (see Sect. \ref{chap:adversarial-behavior-analysis}).
		Although our approach is tool-agnostic, \gls{tamarin}, which is utilized for the prototype, does not provide traces directly.
		However, we leverage \gls{tamarin}'s ability to render trace graphs, not only via \emph{DOT/Graphviz}, but also via experimental \emph{JSON output}.
		We provide a custom parser for the undocumented JSON format enabling structured trace graph generation.\\
		For \emph{Eventlog Conversion}, we first filter out traces that do not depict \gls{sp} invalidation.
		We exploit that trace graphs are structured as a \acrfull{dag}.
		Using the trace graphs of each invalidated lemma we utilize a topological sort to obtain ordered trace steps.
		We apply this to all counterexample traces of the respective \gls{sp} and collect them into sets.\\
		From these sets of ordered steps we extract event-log–relevant information (e.g. activity, affected component).
		Each trace receives a unique ID and incremental time-stamps are assigned by step order and case.
		Finally, event-logs are exported as separate \emph{CSV} files per \gls{sp}.\\
		The \gls{pm} analysis can be performed on these event-logs.
		For this step, we use \gls{prom}~\cite{aalstProMProcessMining2009}, though any \gls{pm} tool suffices.
		In \gls{prom}, we e.g. apply \emph{\gls{idhm}}~\cite{mannhardtHeuristicMiningRevamped2017} and \emph{Convert log to directly follows graph}~\cite{leemansScalableProcessDiscovery2018} modules after converting CSV to \emph{\gls{xes}} and applying filters.\\
		The \gls{roadminer} results provide evaluations including \emph{\glspl{dfg}}, \emph{dependency graphs}, \emph{process trees}, \emph{Petri nets}, and \emph{causal nets}, enabling users to trace attacker behavior leading to \gls{sp} invalidation, i.e. the specific required steps.

	\subsection{Evaluation of Case Study}
		\begin{wrapfigure}{r}{0.6\textwidth}
			\centering
			\vspace*{-2em}
			\includegraphics[width=1\linewidth]{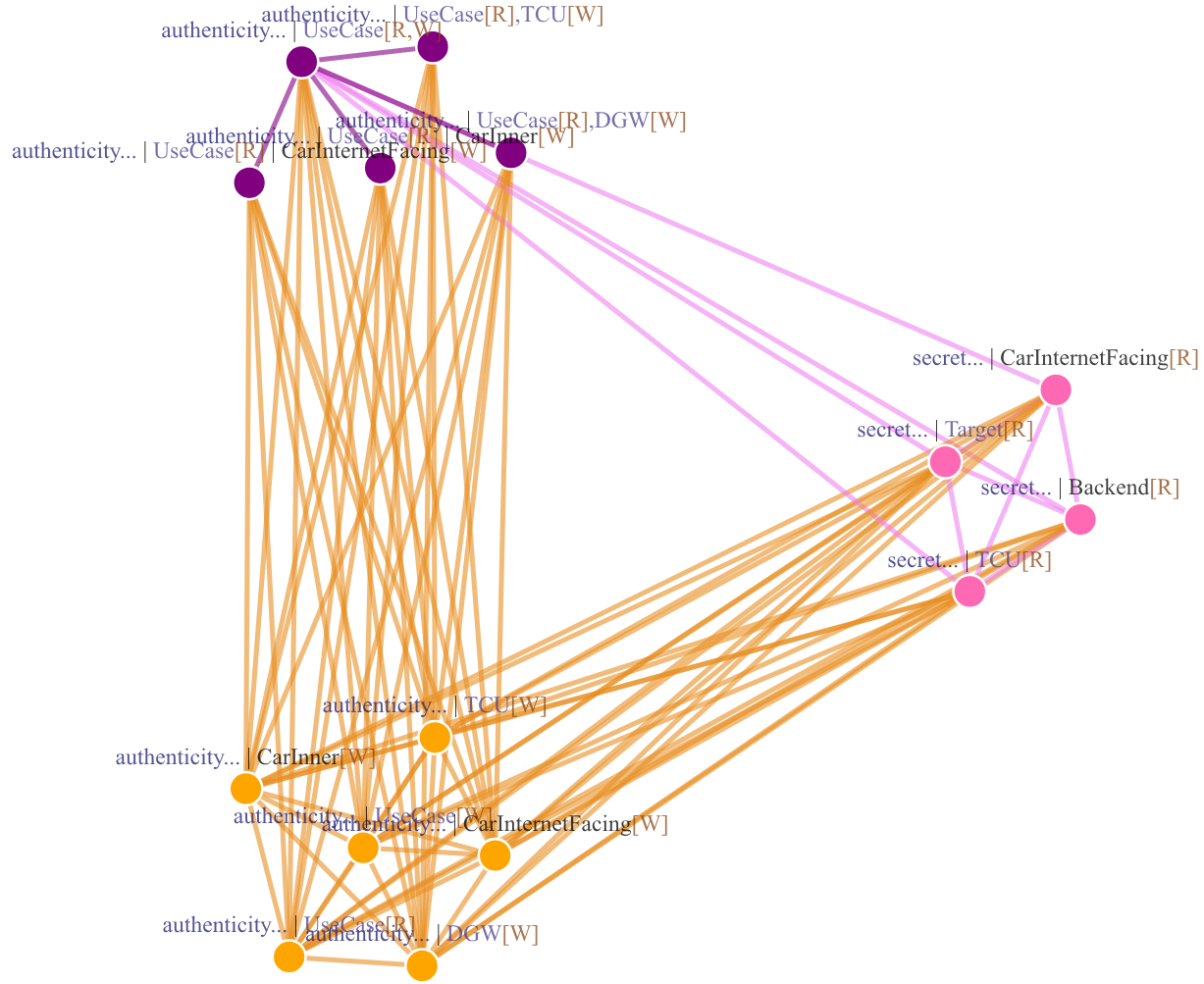}
			\vspace*{-1.25em}
			\caption{Filtered \acrshort{impact} Result-Graph}
			\label{fig:impact-graph-results-secrecy-and-authenticity}
			\vspace*{-2em}
		\end{wrapfigure}
		Our contributions, implemented in the prototypical tools \gls{impact} and \gls{roadminer}, are assessed via a case study on uploading \gls{bms} data to the backend (see Sect. \ref{chap:case-study}).
		We demonstrate their value by evaluating \emph{secrecy of transmitted data} and \emph{authenticity of communication} between the \emph{UseCase} and \emph{Target} (see Sect. \ref{chap:security-properties}) within the case study protocol.\\
		We analyze the results produced by \gls{impact} and presented in the web interface.
		Beyond the comprehensive tabular analysis (see Fig. \ref{fig:impact}), we highlight the concise, accessible presentation of \gls{impact} results (see Fig. \ref{fig:impact-graph-results-secrecy-and-authenticity})\footnote{Results are dynamically highlighted for readability, which the screenshot does not capture.}.
		For the result evaluation, we use the ability to filter \gls{arh} outcomes for \emph{Minimal Compromise Subset} (purple), \emph{Single Point of Failure} (pink), and \emph{Necessary but not Sufficient} (yellow).\footnote{Compromise scenarios regularly verified and invalidated, as well as those labeled \emph{Never responsible for Compromise}, are not of interest here.}%
		The results group into three clusters, reflecting the two verified lemmas (\gls{sp}) and their invalidation by the respective compromises.

		The first cluster covers secrecy invalidation: confidentiality of \emph{UseCase} provided data~${\sim}n$ and \emph{\gls{tcu}}-provided information~${\sim}oN$, composed of \emph{Single Point of Failure} \gls{arh} components.
		For this \gls{sp}, each minimal compromise scenario has a single element, so secrecy can be breached by compromising just one component.
		This invalidation requires only \emph{read compromise}, either of the \glspl{ecu} \emph{\gls{tcu}} or \emph{Target}, or the \emph{CarInternetFacing} or \emph{Backend} domains.
		This is expected in the protocol's context: the \gls{tcu} supplies the additional information for invalidation ${\sim}oN$ only in later stages.
		Thus, only components at or beyond the \gls{tcu} in the protocol execution suffice to breach the property.
		\emph{Write} access is not required for knowledge extraction.

		In contrast to the secrecy \gls{sp}, authenticity (see List. \ref{list:tamarin-lemma-authenticity}) results split into two clusters.
		The first cluster contains single-component elements whose compromise is \emph{necessary but not sufficient} for invalidation.
		Notably, only compromising the \emph{UseCase} is \emph{read}-based; all other related scenarios are \emph{write}-based.
		These appear in the second cluster, where each write-based compromise combined with the UseCase read-compromise forms a \emph{Minimal Compromise Subset} that invalidates authenticity.
		The ability to evaluate multiple \glspl{sp} simultaneously enables comparison of component involvement.
		It is apparent that, aside from the \emph{CarInternetFacing} domain and the \emph{\gls{tcu}} appearing in multi-component compromises for authenticity, the components required for invalidation diverge.
		Furthermore, the common components require read permissions for secrecy but write permissions for authenticity.\\
		To break authenticity, the UseCase must be impersonated through injection of a malicious message.
		This requires a read-based compromise of the UseCase to extract its private key~${\sim}pkUC$ and message injection in the protocol flow before \gls{tcu} processing .

		\begin{wrapfigure}{l}{0.5\textwidth}
			\begin{minipage}{0.5\textwidth}
				\vspace{-2.5em}
				\begin{lstlisting}[caption={Authenticity Lemma}, label={list:tamarin-lemma-authenticity}]
lemma authenticity_of_use_case:
" All useCase target nonce #j .
	EndWithNonce(useCase,target,nonce)@j
	& (not Ex entity #r . PkiPrivatekeyReveal(entity)@r)
		==> Ex #i .
			InitiateTransport(useCase,target,nonce)@i
			& #i < #j "
				\end{lstlisting}
				\vspace{-3em}
			\end{minipage}
		\end{wrapfigure}
		Our extended \gls{arh} analysis with \gls{impact} supports analysis of complex \gls{ana}, comparatively examining component roles in invalidating multiple \glspl{sp} under fine-grained permissions, answering \enquote{\emph{who} is responsible?}.
		For secrecy, the behavior enabled by the compromises that lead to \gls{sp} invalidation, i.e. \enquote{\emph{how} is the \gls{sp} invalidated?}, can still be intuitively inferred.
		This already becomes challenging when evaluating authenticity.

		\begin{figure}[hbt]
			\centering
			\vspace{-1em}
			\includegraphics[width=1\linewidth]{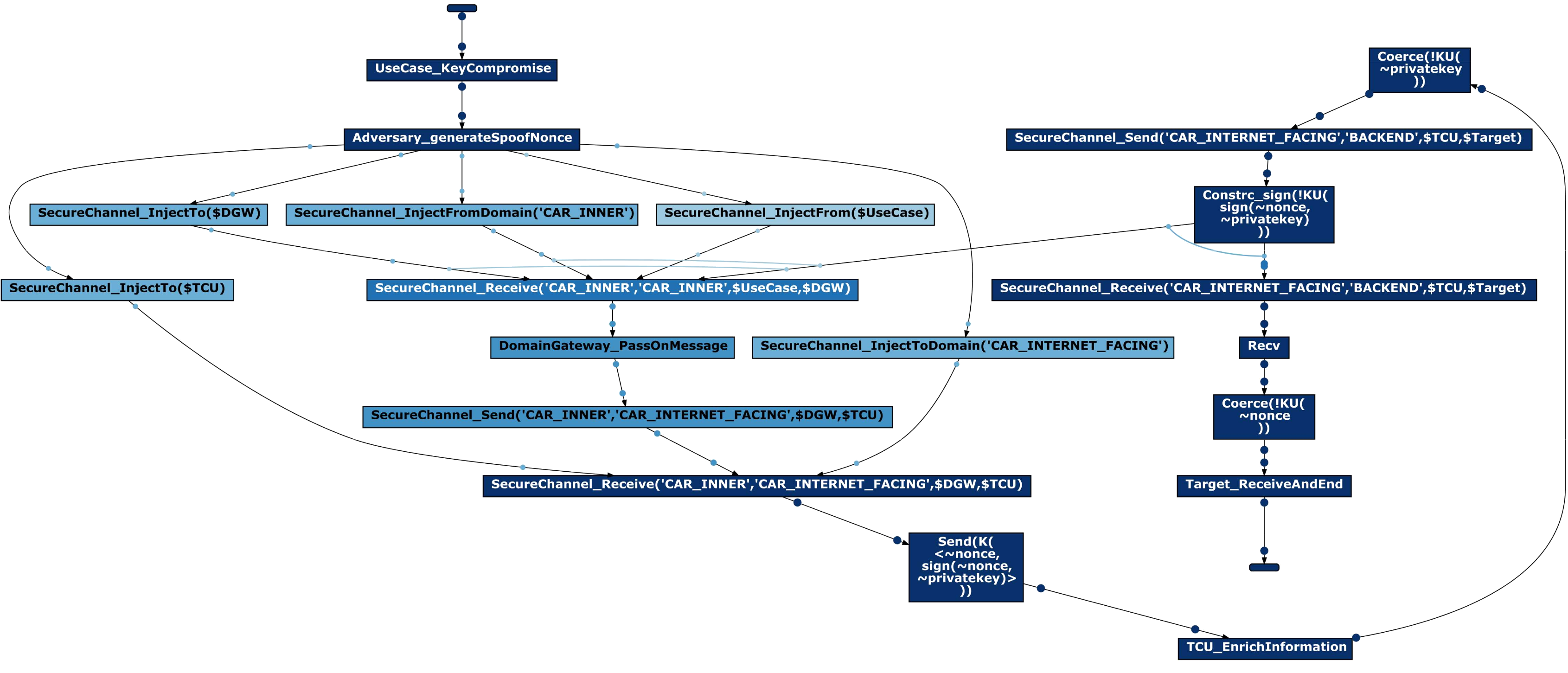}
			\vspace{-2em}
			\caption{Authenticity Violating Causal-Net}
			\vspace{-2.25em}
			\label{fig:filtered-authenticity-idhm-causal-net-flexibleheuristicsminer}
		\end{figure}
		To address this, we extend the analysis with \gls{roadminer} to identify behavior that invalidates message authenticity.
		We use \gls{pm} to capture not only the \enquote{What} but also the \enquote{How} and synthesize adversarial behaviors across all compromise scenarios.
		\gls{roadminer} uses \gls{impact}'s intermediate results as the basis for analysis, converting and aggregating them into synthetic event-logs.
		To ease processing, we provide logfiles in \gls{xes} format.\\
		The event-log was then imported into \gls{pm} software; we used the open-source \gls{prom}.
		We filtered the event-log to improve clarity.
		Steps irrelevant to invalidation (e.g. component initialization and \acrshort{pki} registration) were excluded, assumed established before protocol start and attacker actions.

		\begin{wrapfigure}{l}{0.65\textwidth}
			\centering
			\vspace*{-2em}
			\includegraphics[width=1\linewidth]{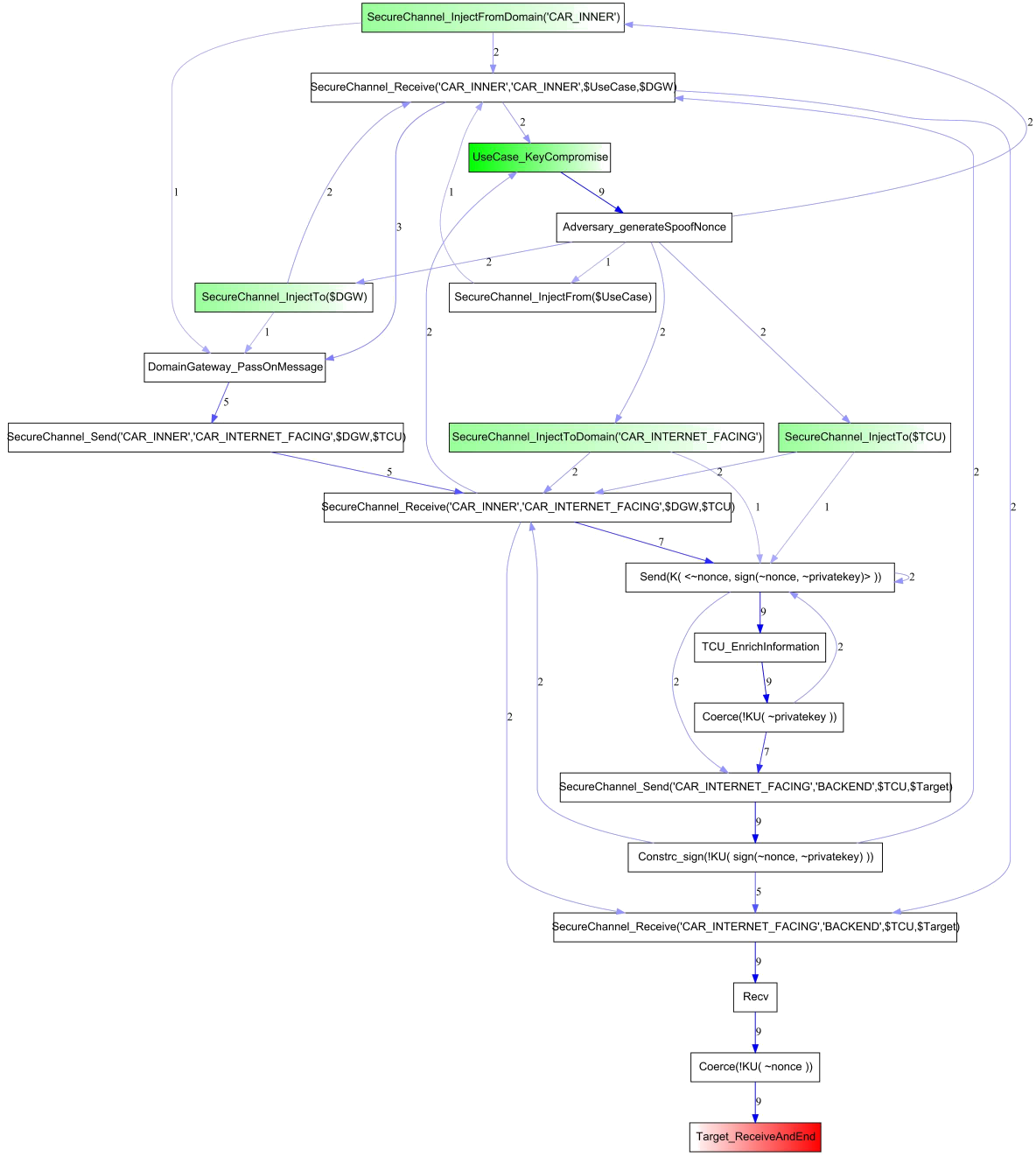}
			\vspace*{-1.75em}
			\caption{Authenticity violating \acrshort{dfg}}
			\label{fig:filtered-authenticity-convert-log-to-directly-follows-model}
			\vspace*{-2.25em}
		\end{wrapfigure}
		Afterward, we analyzed the \emph{authenticity} event-log with the \emph{\gls{idhm}} module~\cite{mannhardtHeuristicMiningRevamped2017} and produced a \emph{Causal net} (see Fig. \ref{fig:filtered-authenticity-idhm-causal-net-flexibleheuristicsminer}) and a \emph{\gls{dfg}} (see Fig. \ref{fig:filtered-authenticity-convert-log-to-directly-follows-model}).
		We condense all adversarial behavior that invalidates the authenticity \gls{sp}, the adversarial and protocol steps of every invalidating compromise scenario trace in the \gls{tamarin} results, into the corresponding graphs via \gls{pm}.
		This higher information density provides a comprehensive overview of all possible attacks and their required steps.
		Both graphs resemble bottom-up \gls{pm} approaches: the \emph{Causal net} uses the \emph{Flexible Heuristic Miner} dependency heuristic and \emph{Cohen's Kappa} as the conditional heuristic with default settings, and the \gls{dfg} is generated with \emph{event name} as the event classifier and \emph{event classes} as the variant.

		Using \gls{fv}-based \gls{arh} analysis, we identify compromised components causing \gls{sp} invalidation.
		The \gls{pm} result graphs (see Fig.~\ref{fig:filtered-authenticity-idhm-causal-net-flexibleheuristicsminer} and Fig.~\ref{fig:filtered-authenticity-convert-log-to-directly-follows-model}) clarify attacker intervention points for breaking authenticity in the protocol:
		Compromising the UseCase private key~${\sim}pkUC$ is the invalidation prerequisite and start of the \gls{pm} graphs.
		This matches the \gls{arh} \enquote{necessary but not sufficient} for UseCase read compromise.
		The adversary generates a spoofed nonce and signs it with the compromised key.
		Consistent with the \gls{arh} analysis, a write compromise of the \gls{tcu}, \acrshort{dgw}, or a vehicular domain (Inner or InternetFacing) lets the adversary inject the spoofed nonce, thus invalidating authenticity.
		This aligns with the \gls{arh} \enquote{Minimal Compromise Subset} results.\\
		The injection must occur before the \gls{tcu} in the protocol run, since it adds additional information and also signs the UseCase nonce.
		Thereafter, the protocol proceeds normally.

	\subsection{Correctness and Limitations of Results}
		Our case study shows that our approach augments \gls{arh} evaluation of invalidation responsibility by illustrating attacker strategies with \gls{roadminer}.
		We extend \gls{arh} analysis to comparative assessment of \glspl{sp} and to attribute invalidation to specific, permission-based compromises.
		The complex analysis required is enabled by our verification-orchestration algorithm.
		We summarize and condense numerous non-machine-readable \gls{tamarin} traces with \gls{pm}, producing clear and analyzable models.
		We provide a concise, machine-analyzable summary of invalidating traces using \gls{pm}.\\
		The diverse results are, aside from minor details, content-identical and reflect actual behavior in the \gls{tamarin} traces.
		Our comparison of \gls{tamarin}-identified traces with the constructed process models supports performance assessment:
		There is no \emph{generalization}; the model captures only attacker behavior in invalidating traces and maintains \emph{precision}; only trace-contained activities are present.
		From a security architect's perspective, we achieve a meaningful balance between \emph{simplicity} (simplifying relations and filtering activities) and \emph{replay fitness} (including all relevant steps as activities).\\
		Using \gls{pm} requires trade-offs for complex relationships (e.g. many invalidating traces) in one graph.
		With effective representations (e.g. \gls{dfg}) and pre-filtering irrelevant steps, density reduces significantly.
		A balance between complexity and simplified relationships must be maintained.
		Given expected partial compromises in automotive systems, focusing only on protecting critical components is insufficient to prevent \gls{sp} invalidation.
		From a security architect's perspective, \gls{pm} graphs help evaluate attacker behavior and prevent invalidating steps.
		Beyond the alignment between compromise scenarios and attacker behavior observed in \gls{pm}, correctness can be verified intuitively:
		First, identify invalidating attacker behavior in the \gls{pm} results.
		Leverage \gls{pm} features (e.g. \glspl{dfg}) to highlight frequent invalidating behaviors.
		Then refine the protocol to prevent the behavior.
		Finally, rerun analysis with \gls{impact} and \gls{roadminer} to confirm the cause for invalidation is no longer present.

	\section{Related Work}\label{chap:related-work}
	Our contribution is a fine-grained model of attacker capabilities for component compromise in automotive network architectures.
	\cite{durrwangAutomationAutomotiveSecurity2021} use a similar approach, using \enquote{attacker privileges} to constrain actions (e.g. read, write) on components such as \glspl{ecu}.
	In contrast to their focus on attack-tree generation, we utilize this approach to define a fine-grained attacker model and its capabilities for security protocol verification.
	We propose \gls{crash-m}, an automotive- and \acrlong{cps}-specific extension of the \gls{dy-m} adversary model that includes component and network-segment compromise and the injection or extraction of entity-internal knowledge.
	While~\cite{rocchettoCPDYExtendingDolevYao2016} extends the \gls{dy-m} for \acrlongpl{cps} (hardware/physical attacks) and~\cite{basinKnowYourEnemy2014} integrates computational aspects, our approach captures properties crucial for modeling adversaries in automotive network architectures.

	Our contribution improves feasibility and performance (runtime and efficiency) in analyzing \glspl{arh}, focusing on security \acrlongpl{hp} and network architectures via security protocol verification.
	In contrast to~\cite{finkbeinerAlgorithmsModelChecking2015}, which verifies \gls{hp} of hardware modules using automata-based algorithms in HyperLTL and HyperCTL$^*$, and~\cite{niessenFindingCounterexamples2024}, which proposes an algorithm for~$\forall\exists$-safety in infinite-state software systems, our work targets security-specific \gls{hp}.

	Our work extracts extended security-property information, including adversarial attack patterns, via \acrfull{pm} of synthetic event-logs from security protocol verification traces.\\
	\Gls{pm} has been applied to security;~\cite{vanderaalstProcessMiningSecurity2005} analyzed audit trails using the~$\alpha$-algorithm to detect violations.
	However, to our knowledge, no prior work uses outputs of formal or security protocol verification tools, focused on \glspl{sp}, as input to \gls{pm}.\\
	\cite{casaluceEnhancingThreatModel2024} use statistical model checking to simulate event-logs and \gls{pm} to build a \enquote{diff} model highlighting discrepancies between expected and observed behavior.
	Their analysis targets real-world physical threat models (e.g. bank robberies) rather than network architectures or communication protocols.\\
	Prior work has explored the reverse direction: combining \gls{pm} with formal methods.
	\cite{vanderaalstProcessMiningVerification2005} employ event logs as input for model checking, specifically leveraging LTL-based techniques.
	\cite{martinelliModelCheckingBased2019} utilized \gls{pm} to derive labeled transition systems.
	Furthermore,~\cite{knupleschEnablingDataAwareCompliance2010} integrate compliance rules into process models for checking and visualizing results directly within the model.

	\section{Conclusion and Outlook}\label{chap:conclusion-and-outlook}
	This section concludes by distilling this work's contributions and positioning them in terms of their significance and limitations.\\
	We propose a novel adversarial model \emph{\gls{crash-m}} \textbf{C1}, representing a strong active attacker tailored to fit and capture the specifics of the automotive domain.
	Consequently, we answered \textbf{RQ1} and established part of the formal foundation for \textbf{C2}.
	Although the attacker model is tailored to automotive use cases due to domain-specific customization, it can be generalized to \acrlongpl{cps} and, more broadly, to network architectures.\\
	We enable impact identification for component compromise \textbf{C2} through improved \gls{arh} analysis, enabling finer-grained properties in more complex \glspl{ana}.
	We achieve this with a novel \gls{arh} verification-orchestration algorithm \textbf{C2.1} that substantially extends \acrfull{arh} analysis capabilities to complex \acrfullpl{ana} and answers \textbf{RQ2}.
	It also enables comparative multi-\acrfull{sp} analysis \textbf{C2.2} (answering \textbf{RQ3}) and a fine-grained adversarial permission model \textbf{C2.3} (answering \textbf{RQ4}).
	This extended \gls{arh} analysis attributes \gls{sp} invalidations to permissions and components (entities, domains) in \glspl{ana} compromise scenarios.
	It reveals which component–permission combinations, as compromise scenarios, invalidate which \gls{sp}.
	A limitation is that it identifies relevant components but not the attacker behavior causing invalidation.\\
	We address this limitation through comprehensive adversarial-behavior analysis \textbf{C3}, answering \textbf{RQ5}.
	Leveraging \gls{arh} verification results, we synthesize event-logs (\textbf{C3.2}) upon which we apply \acrfull{pm} (\textbf{C3.1}).
	This novel approach, bridging \gls{fv} and \gls{pm}, reveals attacker behavior leading to \glspl{sp} invalidations, enabling systematic \gls{ana} hardening.\\
	To demonstrate real-world applicability of our contributions, we verify the security of an example \gls{ana}'s \acrlong{bms} protocol via a case study.
	We operationalize \textbf{C1–C3} with two prototypes, \emph{\acrfull{impact}} and \emph{\acrfull{roadminer}}, comprising \textbf{C4}.
	The case study evaluation confirms substantial benefits for formal security verification of \glspl{ana}.
	However, given the prototype nature of the implementation, the full potential of \acrshort{impact} and \acrshort{roadminer} has not yet been fully realized.

	We outline future research directions that build on our contributions and help remove current limitations.
	Fundamentally, \gls{impact} and \gls{roadminer} can be integrated into a cohesive, fully automated tool, utilizing \gls{pm} libraries, like e.g. PM4Py~\cite{bertiPM4PyProcessMining2023} rather than a GUI tool.
	We have intentionally deferred this to allow qualitative, case-specific log filtering during pre-processing.
	Further implementation is required for more advanced automation.
	This automation is a promising research direction, towards improving design and verification of \glspl{ana} and security protocols.
	Moreover, further insights could be gained through more sophisticated \gls{pm} analyses.
	Rather than running analyses outside the \gls{pm} toolchain, advanced evaluations could be embedded directly in the \gls{pm} workflow, e.g. scoring, categorizing, and identifying attacker behavior patterns.
	Additionally, further exploration of extended \gls{pm} applications is worthwhile.
	For instance, \gls{pm} could support creating general or domain-specific artifacts (e.g. \glspl{tara}), deriving attack trees, and informing component-specific threat and risk assessments.

	\begin{credits}
		\subsubsection{\discintname}
			The authors have no competing interests to declare that are relevant to the content of this article.
	\end{credits}

	\bibliographystyle{splncs04}
	\bibliography{used-references}
\end{document}